\documentclass[a4paper,11pt]{article}
\usepackage{jheppub}
\usepackage[T1]{fontenc}
\usepackage{amsfonts}
\usepackage{amsmath}
\usepackage{amssymb,epsf}
\usepackage{latexsym}
\usepackage{graphicx,epsfig}
\usepackage{amssymb}
\usepackage{subfigure}

\title{Thermodynamics and gauge/gravity duality for Lifshitz black holes in the presence of exponential electrodynamics}
\author[a]{M. Kord Zangeneh}
\author[a]{A. Dehyadegari}
\author[a,b]{A. Sheykhi}
\author[a,b]{M. H. Dehghani}
\affiliation[a]{ Physics Department and Biruni Observatory, College of Sciences, Shiraz University, Shiraz 71454, Iran}
\affiliation[b]{ Research Institute for Astrophysics and Astronomy of Maragha (RIAAM), P.O. Box 55134-441, Maragha, Iran}
\emailAdd{mkzangeneh@shirazu.ac.ir}
\emailAdd{asheykhi@shirazu.ac.ir}
\emailAdd{mhd@shirazu.ac.ir}
\abstract{In this paper, we construct a new class of topological black hole Lifshitz
solutions in the presence of nonlinear exponential electrodynamics for Einstein-dilaton
gravity. We show that the reality of Lifshitz supporting Maxwell matter fields exclude
the negative horizon curvature solutions except for the asymptotic AdS case. Calculating
the conserved and thermodynamical quantities, we obtain a Smarr type formula for the mass and confirm that
thermodynamics first law is satisfied on the black hole horizon. Afterward, we study
the thermal stability of our solutions and figure out the effects of different parameters
on the stability of solutions under thermal perturbations. Next, we apply the gauge/gravity
duality in order to calculate the ratio of shear viscosity to entropy for a three-dimensional
hydrodynamic system by using the pole method. Furthermore, we study the behavior of holographic
conductivity for two-dimensional systems such as graphene. We consider linear Maxwell and
nonlinear exponential electrodynamics separately and disclose the effect of nonlinearity
on holographic conductivity. We indicate that holographic conductivity vanishes for $z>3$
in the case of nonlinear electrodynamics while it does not in the linear Maxwell case. Finally,
we solve perturbative additional field equations numerically and plot the behaviors of
real and imaginary parts of conductivity for asymptotic AdS and Lifshitz cases. We present
experimental results match with our numerical ones.}
\begin{document}
\maketitle
\flushbottom

\section{Introduction}

In the first decades of last century, physics encounters two great
revolutions; one in large scale structures' domain namely special and
general relativities and another in microscopic world i.e. quantum
mechanics. Since quantum mechanics deals with low speed microscopic
particles, physicists quickly thought about a new theory combining (at the
first step) special relativity and quantum mechanics called relativistic
quantum mechanics in order to describe high speed microscopic particles. In
this regime, we continuously encounter production and annihilation of
particles. Therefore, relativistic quantum mechanics was progressed and gave
birth to a more consistent theory i.e. quantum field theory (QFT) where the
Hilbert space is vast enough to include arbitrary number of particles. QFT
indicates impressive theoretical results confirmed up to many significant
digits by experiments in the case of weak interactions where it is allowed
to perform perturbation method \cite{Fey}. However, when coupling is strong,
one should follow other ways in order to compute physical quantities in the
context of QFT.

Fundamental particles physics is not the only place where QFT is employed.
In fact, the situations where one faces non-relativistic particles produced
and annihilated continuously are the other places that QFT methods can be
utilized. The most familiar situations with this behavior are condensed
matter systems. For instance, phonons in a lattice are produced and
annihilated. Also, electrons below or above Fermi level may pass it and
cause a hole production or annihilation. These events resemble the pair
production or annihilation. Another place where QFT methods are useful is
statistical field theory. When we have a system with continuous
configurations, the statistical mechanics of the system is formulated in the
context of statistical field theory. Accordingly, we deal with an
integration in the formula of central quantity of statistical mechanics
namely partition function instead of the usual summation. In the other hand,
the amplitude of the transition between two states in QFT interpreted as the
generating function is calculated through the path integral method and has
very similar formulation to the partition function of statistical field
theory. Hence, the methods of QFT can be effectively applied in these cases.
A very important example of such situation happens near a critical point
where we have a continuous phase transition.

As we mentioned above the common perturbation method does not work for
strong interactions in QFT and hence one should look for another methods to
calculate quantities in this case. One of these methods based on physics of
black holes (mostly thermodynamics of black holes) is gauge/gravity duality
(GGD). The first version of GGD is AdS/CFT correspondence presented by
Maldacena in 1998 \cite{Mald}. GGD connects the QFT (usually gauge field
theories) lives on $n$-dimensional boundary of the spacetime with black hole
solutions in $(n+1)$-dimensional gravity theories of the bulk. This
connection is performed by equality of generating function of QFT ($Z_{QFT}$%
) and the saddle-point approximation of bulk partition function in the case
of classical gravity. Through this, one can obtain the key quantity of QFT
namely generating function. Computation of almost all physical quantities in
QFT relies on generating function $Z_{QFT}$.

The idea of GGD has been applied frequently in order to analyse condensed
matter systems. For instance quantum Hall effect \cite{Hall}, fractional
quantum Hall effect \cite{FQH}, Nernst effect \cite{Ner} and superconductors 
\cite{SupCond,SupCondLif} have been studied by this method. Furthermore,
there are interesting strongly correlated electronic and atomic systems and
also non-relativistic ones which possess Schrodinger symmetry \cite{Schr}.
Nevertheless, the near a critical point dynamics of such systems can be
described by a relativistic conformal field theory or a more subtle scaling
theory respecting the Lifshitz symmetry \cite{Lif}%
\begin{equation}
t\rightarrow \lambda ^{z}t,\text{ \ \ \ \ }\vec{\mathbf{x}}\rightarrow
\lambda \vec{\mathbf{x}}.
\end{equation}%
The spacetime which supports above symmetry on its $r$-infinity boundary is 
\cite{Lif} 
\begin{equation}
ds^{2}=-\frac{r^{2z}}{l^{2z}}dt^{2}+\frac{l^{2}dr^{2}}{r^{2}}+r^{2}d\vec{%
\mathbf{x}}^{2},  \label{lifmet}
\end{equation}%
known as Lifshitz spacetime. In (\ref{lifmet}), $z$ is dynamical critical
exponent. As regards the fact that thermodynamics of black holes plays a
central role in GGD, the literature has encountered increasing interests on
this issue. Solutions and thermodynamics of asymptotic Lifshitz black
objects (usually called Lifshitz solutions) in the presence of massive gauge
fields have been studied in \cite{masssol} and \cite{Deh1} respectively.
Also, thermodynamics of Lifshitz solutions of gravity models containing
higher curvature corrections has been considered under study \cite{hcc}.
Lifshitz solutions with toroidal event horizons and their thermal properties
in the framework of gravity theory coupled to Abelian gauge fields with
negative cosmological constant have been studied in \cite{toplif}. Moreover,
in Einstein-dilaton gravity in the presence of massless gauge fields,
thermal behavior of uncharged \cite{peet} and linearly charged \cite{tario}
Lifshitz black solutions have been investigated. In this context, also
nonlinearly charged Lifshitz solutions with power-law Maxwell field have
been taken under consideration \cite{Deh2}. Furthermore, thermodynamics of
Gauss-Bonnet-dilaton Lifshitz black branes has been discussed in \cite{Deh3}%
. The idea of holography has been applied for systems with Lifshitz symmetry
in different contexts including dilaton gravity from different points of
view \cite{hololif} and behaviors of some quantities such as viscosity \cite%
{vislif} and different kinds of conductivity \cite%
{SupCondLif,CondLif,1106,1211,1311,1402,1508} have been analyzed. The
authors of Ref. \cite{SupCondLif}, constructed an Abelian Higgs model in a
gravity background which is dual to a strongly coupled system at a
Lifshitz-fixed point and finite temperature. They also explored the
conductiviy via holographic techniques. The holographic conductivity in the
presence of an additional term in gauge field action including the Weyl
tensor coupled to Maxwell fied strengths in a domain-wall background whose
near horizon IR geometry is Lifshitz black hole and asymptotic geometry is
AdS has been calculated by performing both memberane paradigm and Kubo's
formula \cite{1106}. Employing the gauge/gravity duality the holographic
superconductors were explored with $z=2$ Lifshitz scaling \cite{1211}. The
effects of Lifshitz dynamical exponent $z$ on the holographic superconductor
models including $s$-wave and $p$-wave models have been disclosed both
numerically and analytically in \cite{1311}. Via holography, superconducting
phase transitions of a system dual to Yang-Mills field coupled to an axion
as probes of black hole with arbitrary Lifshitz scaling have been studied
with $p_{x}+ip_{y}$ condensate \cite{1402}. While the latter phase is known
as unstable in the relativistic case ($z=1$) and absence of axion field, in
the non-relativistic case ($z\neq 1$\textbf{) }with axion field, stability
of it has been studied and behavior of Hall conductivity in
non-superconducting pahase has been studied numerically as a function of
Lifshitz scaling \cite{1402}. Furthermore, an analytic computation for
longtitudinal DC conductivity corresponding to Lifshitz-like fixed points
based on linear response theory has been performed in the presence of chiral
anomalies \cite{1508} and an appropriate holographic set up has been
constructed to calculate the Lifshitz sector of the DC conductivity at
strong coupling and low charge density limit.

It is well-known that the behavior of conductivity have a close relation to
electrodynamics model under consideration. There are several nonlinear
electrodynamics models in addition to linear Maxwell one. The first one of
these models is Born-Infeld electrodynamics presented in 1930's \cite{BI}.
This model which arises in open superstrings and D-branes \cite{BI1,BI2},
removes the divergency of charged particles' self energy. The nonlinearity
of this model is determined by a parameter $\beta $ called nonlinear
parameter. Born-Infeld electrodynamics recovers the linear Maxwell model for
large $\beta $'s. Recently, another nonlinear electrodynamics model has been
introduced which is called exponential electrodynamics \cite{hendiexp}. The
large $\beta $ behavior of this model is similar to Born-Infeld
electrodynamics. The advantage of exponential electrodynamics compared to
Born-Infeld theory is that while it does not fully remove the divergency of
electric field at the source, it does make this divergency much weaker in
comparison with Maxwell field \cite{hendi2,hendi3}. This is more reasonable
compared to the Born-Infeld case, since \textit{near the origin} where $%
r\rightarrow0$, the electric field of a point-like charged particle should
be an increasing function. Besides, it has been observed that the
exponential nonlinear electrodynamics has crucial effects on condensation
and critical temperature of a holographic superconductor \cite{ZPCJ.NN}.
Furthermore, in recent studies on holographic superconductors, it has been
shown that the exponential nonlinear electrodynamics can increase the
critical values of the external magnetic field as the temperature goes to
zero \cite{SheAb}. In this paper, we consider Einstein-dilaton gravity in
the presence of exponential electrodynamics coupled to dilaton field. We
construct exact asymptotically Lifshitz black hole/brane solutions and
discuss the thermal behavior of them. Finally, we apply GGD to obtain
viscosity and conductivity corresponding to our solutions for ($3+1$) and ($%
2+1$)-dimensional systems on boundary respectively.

The layout of this paper is as follows. In Sec. \ref{Thermo}, we first find
the exact asymptotic Lifshitz topological black hole solutions of
Einstein-dilaton gravity in the presence of exponential electrodynamics
coupled to dilaton field. Then, we calculate the conserved and
thermodynamics quantities in order to check the satisfaction of
thermodynamics first law. At the end of Sec. \ref{Thermo}, we disclose the
effects of parameters of model on thermal stability of our solutions. In
Sec. \ref{GGD}, we first apply the GGD through pole method to obtain the
viscosity of a system lives on three-dimensions. In continue of Sec. \ref%
{GGD}, we study the holographic conductivity of two-dimensional systems for
both linear Maxwell and nonlinear exponential elctrodynamics. Finally, we
figure out the behavior of real and imaginary parts of conductivity for
asymptotic AdS and Lifshitz cases. The last section is devoted to present
summary and concluding remarks.

\section{Thermodynamics of Asymptotic Lifshitz Solutions}

\label{Thermo} In this section, we first find asymptotic Lifsthitz
solutions. Then, we study thermodynamics of our solutions and show that the
first law of thermodynamics is satisfied on the horizon. Finally, we
investigate thermal stability of the solutions in both canonical and
grand-canonical ensembles.

\subsection{Action and Higher-Dimensional Solutions}

The action of ($n+1$)-dimensional ($n\geq 3$) Einstein-dilaton gravity
coupled to a nonlinear electrodynamics and two linear Maxwell fields can be
written as 
\begin{gather}
S=\int_{\mathcal{M}}d^{n+1}x\mathcal{L},  \label{Act} \\
\mathcal{L}=\frac{\sqrt{-g}}{16\pi }\left( \mathcal{R}-\frac{4}{n-1}(\nabla
\Phi )^{2}-2\Lambda +L(F,\Phi )-\sum\limits_{i=1}^{2}e^{-4/(n-1)\lambda
_{i}\Phi }H_{i}\right) ,  \notag
\end{gather}%
where $\mathcal{R}$ and $\Phi $ are Ricci scalar and dilaton scalar field,
respectively. Here $F=F^{\mu \nu }F_{\mu \nu }$ and $H_{i}=(H_{i})_{\mu \nu
}(H_{i})^{\mu \nu }$, where $F_{\mu \nu }=\partial _{\lbrack \mu }A_{\nu ]}$%
, $(H_{i})_{\mu \nu }=\partial _{\lbrack \mu }(B_{i})_{\nu ]}$, and $A_{\nu
} $ and $(B_{i})_{\nu }$ are electromagnetic vector potentials. $\lambda
_{i} $ and $\Lambda $ are some constants and $L(F,\Phi )$ is the Lagrangian
of nonlinear electrodynamic matter source. Varying the action (\ref{Act})
with respect to the metric $g_{\mu \nu }$, scalar field $\Phi $, and
electromagnetic vector potentials $A_{\nu }$ and $(B_{i})_{\nu }$, leads to
the following equations of motion, 
\begin{eqnarray}
\mathcal{R}_{\mu \nu } &=&\frac{g_{\mu \nu }}{n-1}\left\{ 2\Lambda
+2L_{F}F-L(F,\Phi )-\sum\limits_{i=1}^{2}H_{i}e^{-4\lambda _{i}\Phi
/(n-1)}\right\}  \notag \\
&+&\frac{4}{n-1}\partial _{\mu }\Phi \partial _{\nu }\Phi -2L_{F}F_{\mu
\lambda }F_{\nu }^{\text{ \ }\lambda }+2\sum\limits_{i=1}^{2}e^{-4\lambda
_{i}\Phi /(n-1)}\left( H_{i}\right) _{\mu \lambda }\left( H_{i}\right) _{\nu
}^{\text{ \ }\lambda },  \label{FE1}
\end{eqnarray}%
\begin{eqnarray}
&&\nabla ^{2}\Phi +\frac{n-1}{8}L_{\Phi }+\sum\limits_{i=1}^{2}\frac{{%
\lambda _{i}}}{2}e^{-{4\lambda _{i}\Phi }/({n-1})}H_{i}=0,  \label{FE2} \\
&&\triangledown _{\mu }\left( L_{F}F^{\mu \nu }\right) =0,  \label{FE3} \\
&&\triangledown _{\mu }\left( e^{-{4\lambda _{i}\Phi }/({n-1})}\left(
H_{i}\right) ^{\mu \nu }\right) =0,  \label{FE4}
\end{eqnarray}%
where we have used the convention $X_{Y}=\partial X/\partial Y$. In this
paper, we intend to consider exponential nonlinear electrodynamics \cite%
{hendiexp}. The Lagrangian of such type of nonlinear electrodynamics coupled
to the dilaton field in higher dimensions can be written as \cite%
{SheyHaj,SheyKaz} 
\begin{equation}
L(F,\Phi )=4\beta ^{2}e^{4\lambda \Phi /(n-1)}\left[ \exp \left( \frac{%
-e^{-8\lambda \Phi /(n-1)}F}{4\beta ^{2}}\right) -1\right] .  \label{ELag}
\end{equation}%
The behavior of exponential nonlinear electrodynamics coupled to dilaton
field (END) for large $\beta $, namely 
\begin{equation}
\lim_{\beta \rightarrow \infty }L(F,\Phi )=-e^{-4\lambda \Phi /(n-1)}F+\frac{%
e^{-12\lambda \Phi /(n-1)}F^{2}}{8\beta ^{2}}+O\left( \frac{1}{\beta ^{4}}%
\right) ,  \label{Larbeta}
\end{equation}%
is similar to large $\beta $ behavior of the Born-Infeld electrodynamics
coupled to dilaton field (BID)%
\begin{equation}
\lim_{\beta \rightarrow \infty }L_{BI}(F,\Phi )=-e^{-4\lambda \Phi /(n-1)}F+%
\frac{e^{-12\lambda \Phi /(n-1)}F^{2}}{8\beta ^{2}}+O\left( \frac{1}{\beta
^{4}}\right) ,
\end{equation}%
where%
\begin{equation*}
L_{BI}(F,\Phi )=4\beta ^{2}e^{4\lambda \Phi /(n-1)}\left[ 1-\sqrt{1+\frac{%
e^{-8\lambda \Phi /(n-1)}F}{2\beta ^{2}}}\right] .
\end{equation*}%
This similarity makes the form chosen for END justifiable and also implies
that one can consider both END and BID as the Lagrangian of nonlinear
electrodynamics coupled to dilaton field.

\begin{figure*}[t]
\centering{%
\subfigure[$k=0$, $n=4$]{
   \label{fig1a}\includegraphics[width=.46\textwidth]{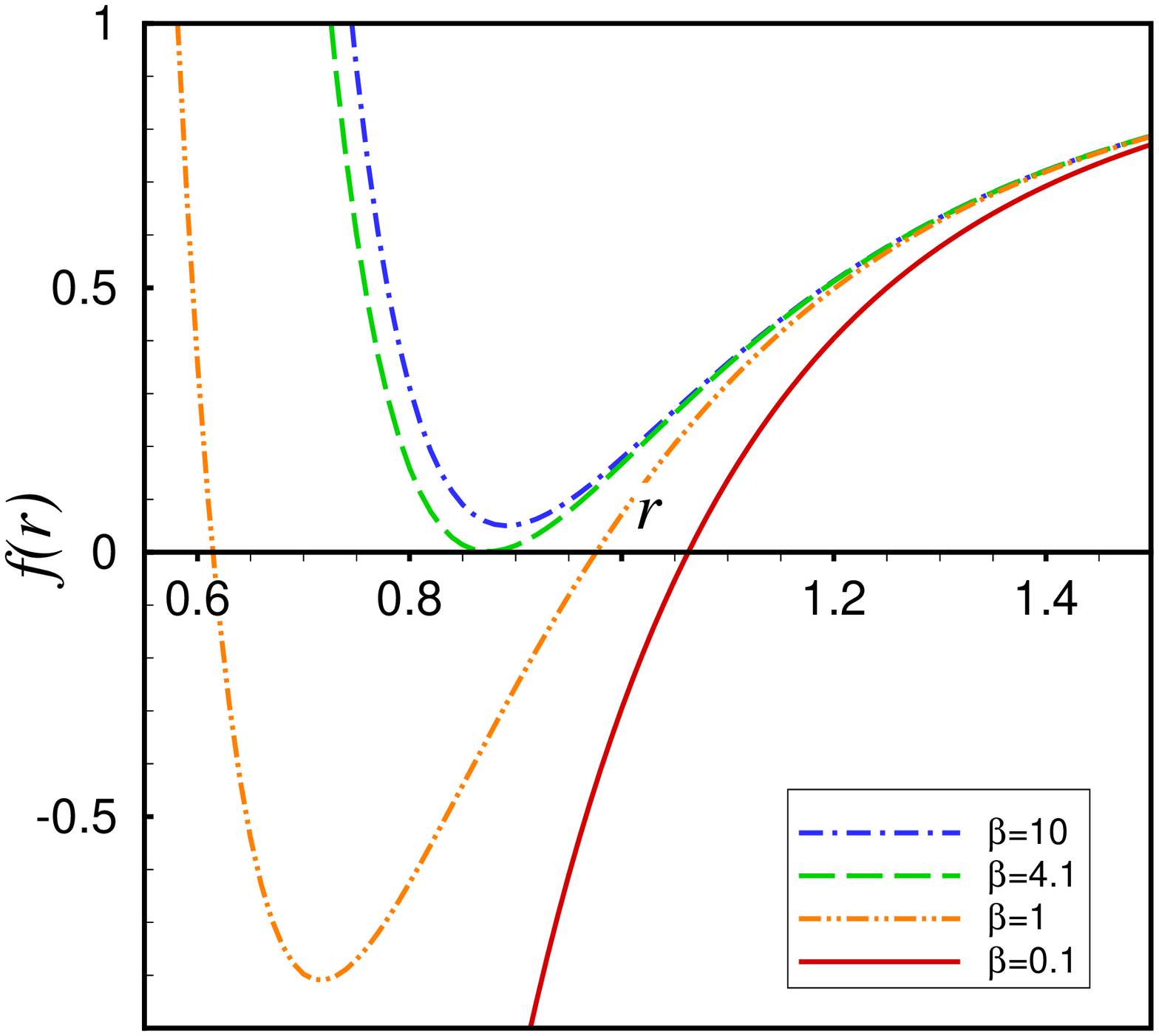}\qquad}} 
\subfigure[$k=1$, $n=5$]{
   \label{fig1b}\includegraphics[width=.46\textwidth]{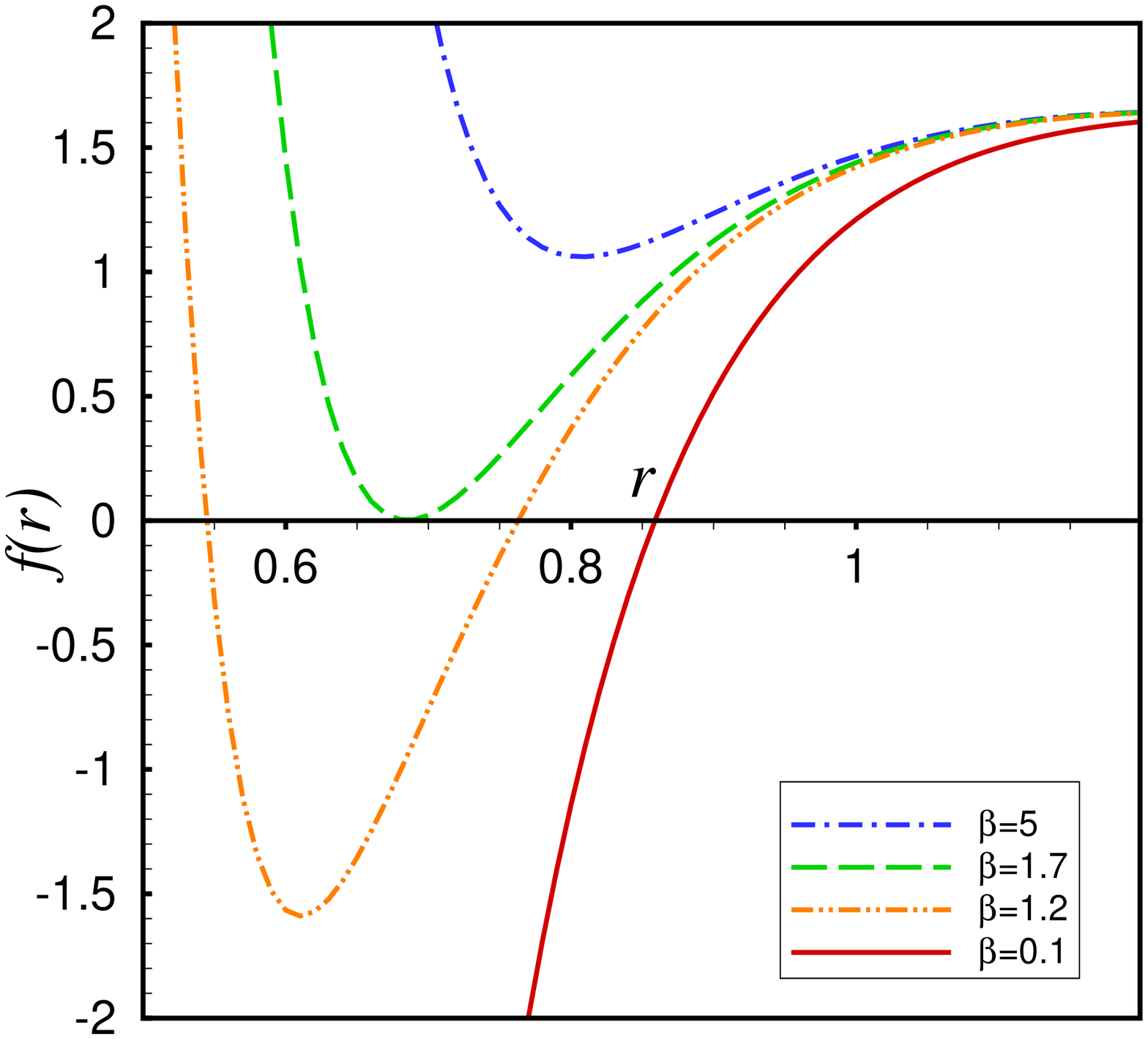}}
\caption{The behavior of $f(r)$ versus $r$ for $l=1.5$, $b=0.3$, $q=1.7$, $%
z=1.5$ and $m=1.6$.}
\label{fig1}
\end{figure*}

In order to construct asymptotic Lifshitz topological black hole solutions
in higher dimensions, we use the metric 
\begin{equation}
ds^{2}=-\frac{r^{2z}f(r)}{l^{2z}}dt^{2}+{\frac{l^{2}dr^{2}}{r^{2}f(r)}}%
+r^{2}d\Omega _{n-1}^{2},  \label{met}
\end{equation}%
where $z\geq 1$ is dynamical critical exponent, $d\Omega _{n-1}^{2}$ is an ($%
n-1$)-dimensional hypersurface with constant curvature $(n-1)(n-2)k$ and
volume $\omega _{n-1}$ and $f(r)\rightarrow 1$ as $r\rightarrow \infty $.
Lifshitz dilaton topological black hole solutions in the presence of linear
Maxwell and nonlinear power-law Maxwell fields have been studied in \cite%
{tario} and \cite{kzsd}, respectively. The first term of (\ref{Larbeta})
recovers the linear case of electrodynamic Lagrangian of \cite{kzsd}.
Therefore, we expect that our solutions recover the results of this case.
Using the metric (\ref{met}), one can immediately integrate (\ref{FE3}) and (%
\ref{FE4}) and find 
\begin{equation}
F_{rt}=\frac{qe^{4\lambda \Phi /(n-1)}}{r^{n-z}}\exp \left[ -\frac{1}{2}%
L_{W}\left( \varrho \right) \right] ,  \label{Frt}
\end{equation}%
\begin{equation}
\left( H_{i}\right) _{rt}=\frac{q_{i}e^{4\lambda _{i}\Phi /(n-1)}}{r^{n-z}},
\label{Hrt}
\end{equation}%
where $\varrho \equiv q^{2}l^{2z-2}/(\beta ^{2}r^{2n-2})$ and $L_{W}(x)=%
\mathrm{Lambert}W(x)$ is the Lambert function which satisfies the identity 
\cite{abram} 
\begin{equation}
L_{W}(x)e^{L_{W}(x)}=x,
\end{equation}%
and has the series expansion 
\begin{equation}
L_{W}(x)=x-x^{2}+\frac{3}{2}x^{3}-\frac{8}{3}x^{4}+\cdots ,  \label{serexp}
\end{equation}%
which converges provided $\left\vert x\right\vert <1$. For $\beta
\rightarrow \infty $, $F_{rt}$ given in Eq. (\ref{Frt}) reduces to the
linear Maxwell case \cite{kzsd} 
\begin{equation}
\lim_{\beta \rightarrow \infty }F_{rt}=\frac{qe^{4\lambda \Phi /(n-1)}}{%
r^{n-z}}-\frac{q^{3}l^{2z-2}e^{4\lambda \Phi /(n-1)}}{2r^{3n-z-2}\beta ^{2}}%
+O\left( \frac{1}{\beta ^{4}}\right) .
\end{equation}%
Substituting Eqs. (\ref{Frt}) and (\ref{Hrt}) into Eqs. (\ref{FE1}) and (\ref%
{FE2}), one can find the following field equations 
\begin{gather}
\frac{(n-1)^{2}rf^{\prime }+(n-1)^{2}nf+4r^{2}f\Phi ^{\prime 2}}{2(n-1)l^{2}}%
+\Lambda -\frac{(n-1)(n-2)k}{2r^{2}}  \notag \\
+\sum\limits_{i=1}^{2}\frac{q_{i}^{2}e^{4\lambda _{i}\Phi /(n-1)}}{%
l^{2(1-z)}r^{2(n-1)}}+4\beta ^{2}e^{4\lambda \Phi /(n-1)}\Theta =0,
\label{fe1}
\end{gather}

\begin{gather}
\frac{(n-1)^{2}rf^{\prime }+(n-1)^{2}(n-2)f+2(n-1)^{2}fz-4r^{2}f\Phi
^{\prime 2}}{2(n-1)l^{2}}+\Lambda -\frac{(n-1)(n-2)k}{2r^{2}}  \notag \\
+\sum\limits_{i=1}^{2}\frac{q_{i}^{2}e^{4\lambda _{i}\Phi /(n-1)}}{%
l^{2(1-z)}r^{2(n-1)}}+4\beta ^{2}e^{4\lambda \Phi /(n-1)}\Theta =0,
\end{gather}

\begin{gather}
\frac{r^{2}f^{\prime \prime }+(2n+3z-3)rf^{\prime }+\frac{4}{n-1}r^{2}f\Phi
^{\prime 2}+(2z^{2}+2(n-2)z+(n-1)(n-2))f}{2l^{2}}+\Lambda  \notag \\
-\frac{(n-3)(n-2)k}{2r^{2}}-\sum\limits_{i=1}^{2}\frac{q_{i}^{2}e^{4\lambda
_{i}\Phi /(n-1)}}{l^{2(1-z)}r^{2(n-1)}}+4\beta ^{2}e^{4\lambda \Phi
/(n-1)}\Psi =0,
\end{gather}

\begin{equation}
\frac{r^{2}f^{\prime }\Phi ^{\prime }+\Phi ^{\prime }frz+nrf\Phi ^{\prime
}+fr^{2}\Phi ^{\prime \prime }}{l^{2}}-\sum\limits_{i=1}^{2}\frac{%
q_{i}^{2}\lambda _{i}e^{4\lambda _{i}\Phi /(n-1)}}{l^{2(1-z)}r^{2(n-1)}}%
-4\beta ^{2}\lambda e^{4\lambda \Phi /(n-1)}\Theta =0,  \label{fe4}
\end{equation}%
where

\begin{equation*}
\Theta =\frac{1}{2}+\frac{ql^{z-1}}{2\beta r^{n-1}}\left( \sqrt{%
L_{W}(\varrho )}-\frac{1}{\sqrt{L_{W}(\varrho )}}\right) ,
\end{equation*}%
and

\begin{equation*}
\Psi =\frac{1}{2}-\frac{ql^{z-1}}{2\beta r^{n-1}\sqrt{L_{W}(\varrho )}}.
\end{equation*}%
The solutions of Eqs. (\ref{fe1})-(\ref{fe4}) can be obtained as 
\begin{equation}
{\Phi (r)}{=}\frac{(n-1)\sqrt{z-1}}{2}{\ln \left( \frac{r}{b}\right) ,}
\label{Phi}
\end{equation}%
\begin{eqnarray}
f(r) &=&1-\frac{m}{r^{n+z-1}}+\frac{(n-2)^{2}kl^{2}}{(n+z-3)^{2}r^{2}} 
\notag \\
&&-\frac{8\beta ^{2}l^{2}b^{2z-2}}{(n-1)(n-z+1)r^{2z-2}}\left\{ \frac{1}{2}+%
\frac{ql^{z-1}(n-z+1)}{2\beta }r^{z-n-1}\int r^{1-z}\left( \sqrt{%
L_{W}(\varrho )}-\frac{1}{\sqrt{L_{W}(\varrho )}}\right) dr\right\} ,  \notag
\\
&&  \label{f}
\end{eqnarray}%
where $m$ is a parameter which is related to the total mass of black hole.
The above solutions will fully satisfy the field equations (\ref{fe1})-(\ref%
{fe4}), provided 
\begin{gather}
\lambda =-\sqrt{z-1},\text{ }\lambda _{1}=\frac{n-1}{\sqrt{z-1}},\text{ }%
\lambda _{2}=\frac{n-2}{\sqrt{z-1}},  \notag \\
q_{1}^{2}=\frac{(n+z-1)(z-1)b^{2(n-1)}}{2l^{2z}},  \notag \\
q_{2}^{2}=\frac{k(n-1)(n-2)(z-1)b^{2(n-2)}}{2(z+n-3)l^{2(z-1)}},  \notag \\
\Lambda =-\frac{(n+z-1)(n+z-2)}{2l^{2}}.  \label{constants}
\end{gather}%
Since in the case of $k=-1$, $q_{2}$ is imaginary (except for $z=1$), we
exclude this case and focus on the black hole ($k=1$) and black brane ($k=0$%
) solutions in the remaining part of this paper. It is notable to mention
that althogh at first glance it seems that constants (\ref{constants})
diverge for $z=1$, one should note that $\lambda _{i}\Phi $\ is finite for
this case and $H_{i}=0$. Hence, for $z=1$, action (\ref{Act}) reduces to
(A)dS action in the presence of a nonlinear electrodynamics field. One can
perform the integration in Eq. (\ref{f}) by using MATHEMATICA software. One
obtains 
\begin{eqnarray}
f(r) &=&1-\frac{m}{r^{n+z-1}}+\frac{(n-2)^{2}kl^{2}}{(n+z-3)^{2}r^{2}} 
\notag \\
&&-\frac{4\beta ^{2}l^{2}b^{2z-2}}{(n-1)(n-z+1)r^{2z-2}}+\frac{8\beta
^{2}l^{2}b^{2z-2}(2n-2)^{\left( z-5n+3\right) /\left( 2n-2\right) }}{%
(z-2)^{(z+3n-5)/(2n-2)}r^{n+z-1}}\left( \frac{q^{2}l^{2z-2}}{\beta ^{2}}%
\right) ^{\left( n-z+1\right) /\left( 2n-2\right) }  \notag \\
&&\times \left\{ 4(n-1)^{2}\left[ \Gamma \left( \frac{3n+z-5}{2n-2},\frac{2-z%
}{2n-2}L_{W}(\varrho )\right) -\Gamma \left( \frac{3n+z-5}{2n-2}\right) %
\right] \right.  \notag \\
&&\left. -(z-2)^{2}\left[ \Gamma \left( \frac{z-n-1}{2n-2},\frac{2-z}{2n-2}%
L_{W}(\varrho )\right) -\Gamma \left( \frac{z-n-1}{2n-2}\right) \right]
\right\} ,  \label{ff}
\end{eqnarray}%
where $\Gamma (x,y)$ and $\Gamma (x)$ are gamma functions and are related to
each other with the relation 
\begin{equation}
\Gamma (x,y)=\Gamma (x)-\frac{y^{x}}{x}\mathbf{F}(x,1+x,-y),
\label{relation}
\end{equation}%
where $\mathbf{F}(a,b,c)$ is the hypergeometric function \cite{abram}. At
the first look, it seems that the function (\ref{ff}) diverges for $z=2$.
However, the factor $(z-2)$ in denominator is removed when one uses Eq. (\ref%
{relation}). Indeed, one can reexpress (\ref{ff}) by using relation (\ref%
{relation}) in terms of the hypergeometric functions as 
\begin{eqnarray}
f(r) &=&1-\frac{m}{r^{n+z-1}}+\frac{(n-2)^{2}kl^{2}}{(n+z-3)^{2}r^{2}} 
\notag \\
&&-\frac{4\beta ^{2}l^{2}b^{2z-2}}{(n-1)(n-z+1)r^{2z-2}}+\frac{4\beta
^{2}l^{2}b^{2z-2}}{(n-1)r^{n+z-1}}\left( \frac{q^{2}l^{2z-2}}{\beta
^{2}L_{W}(\varrho )}\right) ^{\left( n-z+1\right) /\left( 2n-2\right) } 
\notag \\
&&\times \left\{ \frac{L_{W}^{2}(\varrho )}{3n+z-5}\mathbf{F}\left( \frac{%
3n+z-5}{2n-2},\frac{5n+z-7}{2n-2},\frac{z-2}{2n-2}L_{W}(\varrho )\right)
\right.  \notag \\
&&\left. +\frac{1}{n-z+1}\mathbf{F}\left( \frac{z-n-1}{2n-2},\frac{z+n-3}{%
2n-2},\frac{z-2}{2n-2}L_{W}(\varrho )\right) \right\} .  \label{ff1}
\end{eqnarray}%
Using the fact that $f(r_{+})=0$ where $r_{+}$ is the outermost event
horizon, one can obtain%
\begin{equation}
{m(r_{+})}=r_{+}^{n+z-1}+\frac{(n-2)^{2}kl^{2}}{(n+z-3)^{2}r_{+}^{3-n-z}}-%
\frac{8\beta ^{2}l^{2}b^{2z-2}\Xi }{(n-1)(n-z+1)r_{+}^{z-n-1}}  \label{mr+}
\end{equation}%
where 
\begin{eqnarray}
\Xi &=&\frac{1}{2}-\frac{(n-z+1)}{2r_{+}^{n-z+1}}\left( \frac{q^{2}l^{2z-2}}{%
\beta ^{2}L_{W}\left( \varrho _{+}\right) }\right) ^{\left( n-z+1\right)
/\left( 2n-2\right) }  \notag \\
&&\times \left\{ \frac{L_{W}^{2}\left( \varrho _{+}\right) }{3n+z-5}\mathbf{F%
}\left( \frac{3n+z-5}{2n-2},\frac{5n+z-7}{2n-2},\frac{z-2}{2n-2}L_{W}\left(
\varrho _{+}\right) \right) \right.  \notag \\
&&\left. +\frac{1}{n-z+1}\mathbf{F}\left( \frac{z-n-1}{2n-2},\frac{z+n-3}{%
2n-2},\frac{z-2}{2n-2}L_{W}\left( \varrho _{+}\right) \right) \right\} ,
\end{eqnarray}%
and $\varrho _{+}=q^{2}l^{2z-2}/(\beta ^{2}r_{+}^{2n-2})$. The large $\beta $
limit of $f(r)$ can be obtained by using the series expansion (\ref{serexp}%
). The result is 
\begin{eqnarray}
f(r) &=&1-\frac{m}{r^{n+z-1}}+\frac{(n-2)^{2}kl^{2}}{(n+z-3)^{2}r^{2}} 
\notag \\
&&+\frac{2q^{2}b^{2z-2}l^{2z}}{\left( n-1\right) \left( n+z-3\right)
r^{2n+2z-4}}-\frac{q^{4}b^{2z-2}l^{4z-2}}{2\left( n-1\right) \left(
3n+z-5\right) \beta ^{2}r^{4n+2z-6}}+O\left( \frac{1}{\beta ^{4}}\right) . 
\notag \\
&&  \label{flarbeta}
\end{eqnarray}%
As we mentioned before, for $\beta \rightarrow \infty $ the NED Lagrangian
reduces to linear Maxwell Lagrangian. Therefore, we expect that the large $%
\beta $ limit of $f(r)$ recover the metric function of topological Lifshitz
black holes in Maxwell theory \cite{kzsd}. One should note that this is
indeed the case and Eq. (\ref{flarbeta}) recover the linear case of $f(r)$
presented in \cite{kzsd} when $\beta \rightarrow \infty $. The behavior of $%
f(r)$ is depicted in Figs. \ref{fig1a} and \ref{fig1b} for different values
of $\beta $. Figs. \ref{fig1a} and \ref{fig1b} correspond to black brane ($%
k=0$) and black hole ($k=1$) solutions respectively. 
Inserting Eqs. (\ref{Phi}) and (\ref{constants}) into (\ref{Frt}), one finds%
\begin{equation}
F_{rt}=\frac{qb^{2z-2}}{r^{n+z-2}}\exp \left[ -\frac{1}{2}L_{W}\left(
\varrho \right) \right] .
\end{equation}%
The fact that our solutions are static implies $A_{t}=A_{t}\left( r\right) $%
. Hence, we can obtain the gauge potential using relation $A_{t}\left(
r\right) =\int F_{rt}dr$ as%
\begin{eqnarray}
A_{t} &=&\mu +\frac{qb^{2z-2}(2n-2)^{(z-n-1)/(2n-2)}}{%
(2-z)^{(-5+3n+z)/(2n-2)}}\left( \frac{q^{2}l^{2z-2}}{\beta ^{2}}\right)
^{\left( 3-n-z\right) /\left( 2n-2\right) }  \notag \\
&&\times \left\{ 2(n-1)\left[ \Gamma \left( \frac{3n+z-5}{2n-2},\frac{2-z}{%
2n-2}L_{W}\left( \varrho \right) \right) -\Gamma \left( \frac{3n+z-5}{2n-2}%
\right) \right] \right.  \notag \\
&&\left. -(z-2)\left[ \Gamma \left( \frac{z+n-3}{2n-2},\frac{2-z}{2n-2}%
L_{W}\left( \varrho \right) \right) -\Gamma \left( \frac{z+n-3}{2n-2}\right) %
\right] \right\} .  \label{At}
\end{eqnarray}%
It is remarkable to note that since from one side $L_{W}\left( \varrho
\right) \rightarrow 0$ as $r\rightarrow \infty $ and from another side $%
\Gamma \left( x,0\right) =\Gamma \left( x\right) $, one can easily check
that $A_{t}$ is finite at infinity and reduces to the constant $\mu $.
Although (\ref{At}) seems divergent for $z=2$, we can remove the factor $%
(z-2)$ in denominator by using (\ref{relation}) and restate $A_{t}$ as%
\begin{eqnarray}
A_{t} &=&\mu -qb^{2z-2}\left( \frac{q^{2}l^{2z-2}}{\beta ^{2}L_{W}\left(
\varrho \right) }\right) ^{\left( 3-n-z\right) /\left( 2n-2\right) }  \notag
\\
&&\times \left\{ \frac{L_{W}\left( \varrho \right) }{3n+z-5}\mathbf{F}\left( 
\frac{3n+z-5}{2n-2},\frac{5n+z-7}{2n-2},\frac{z-2}{2n-2}L_{W}\left( \varrho
\right) \right) \right.  \notag \\
&&\left. +\frac{1}{n+z-3}\mathbf{F}\left( \frac{n+z-3}{2n-2},\frac{3n+z-5}{%
2n-2},\frac{z-2}{2n-2}L_{W}\left( \varrho \right) \right) \right\} .
\label{At1}
\end{eqnarray}%
We require that $A_{t}$ vanishes at horizon $r=r_{+}$. Thus, $\mu $ can be
calculated as%
\begin{eqnarray}
\mu &=&qb^{2z-2}\left( \frac{q^{2}l^{2z-2}}{\beta ^{2}L_{W}\left( \varrho
_{+}\right) }\right) ^{\left( 3-n-z\right) /\left( 2n-2\right) }  \notag \\
&&\times \left\{ \frac{L_{W}\left( \varrho _{+}\right) }{3n+z-5}\mathbf{F}%
\left( \frac{3n+z-5}{2n-2},\frac{5n+z-7}{2n-2},\frac{z-2}{2n-2}L_{W}\left(
\varrho _{+}\right) \right) \right.  \notag \\
&&\left. +\frac{1}{n+z-3}\mathbf{F}\left( \frac{n+z-3}{2n-2},\frac{3n+z-5}{%
2n-2},\frac{z-2}{2n-2}L_{W}\left( \varrho _{+}\right) \right) \right\} .
\end{eqnarray}%
In the next subsection we study thermodynamics of the solutions by computing
thermodynamic and conserved quantities.

\subsection{Thermodynamics of Lifshitz black holes/branes}

In this section, we study thermodynamics of lifshitz black hole and brane
solutions. Since, the mass is a fundamental quantity in studying the
thermodynamics, we start with its calculation. Using the modified
subtraction method of Brown and York (BY) \cite{modBY}, one can find mass
per unit volume $\omega _{n-1}$ as (see Ref. \cite{kzsd}) 
\begin{equation}
M=\frac{(n-1)m}{16\pi l^{z+1}},  \label{Mass}
\end{equation}%
where $m$ is the geometrical mass given by (\ref{mr+}) in terms of outer
horizon radius $r_{+}$. Using the Gauss law, one can also compute the charge
as 
\begin{equation}
Q=\frac{\,{1}}{4\pi }\int r^{n-1}L_{F}F_{\mu \nu }n^{\mu }u^{\nu }d{\Omega },
\label{chdef}
\end{equation}%
where $n^{\mu }$ and $u^{\nu }$ are the unit spacelike and timelike normals
to a sphere of radius $r$ given as 
\begin{equation*}
n^{\mu }=\frac{1}{\sqrt{-g_{tt}}}dt=\frac{l^{z}}{r^{z}\sqrt{f(r)}}dt,\text{
\ \ \ \ }u^{\nu }=\frac{1}{\sqrt{g_{rr}}}dr=\frac{r\sqrt{f(r)}}{l}dr.
\end{equation*}%
Therefore, the charge per unit volume $\omega _{n-1}$ is 
\begin{equation}
Q=\frac{ql^{z-1}}{4\pi }.  \label{charge}
\end{equation}%
The electric potential $U$ is defined as%
\begin{equation}
U=A_{\mu }\chi ^{\mu }\left\vert _{r\rightarrow \infty }-A_{\mu }\chi ^{\mu
}\right\vert _{r=r_{+}},  \label{Pot}
\end{equation}%
where $\chi =\partial _{t}$ is the null generator of the horizon. Therefore,
it is a matter of calculation to obtain $U$ by using (\ref{At1}) as 
\begin{eqnarray}
U &=&qb^{2z-2}\left( \frac{q^{2}l^{2z-2}}{\beta ^{2}L_{W}\left( \varrho
_{+}\right) }\right) ^{\left( 3-n-z\right) /\left( 2n-2\right) }  \notag \\
&&\times \left\{ \frac{L_{W}\left( \varrho _{+}\right) }{3n+z-5}\mathbf{F}%
\left( \frac{3n+z-5}{2n-2},\frac{5n+z-7}{2n-2},\frac{z-2}{2n-2}L_{W}\left(
\varrho _{+}\right) \right) \right.  \notag \\
&&\left. +\frac{1}{n+z-3}\mathbf{F}\left( \frac{n+z-3}{2n-2},\frac{3n+z-5}{%
2n-2},\frac{z-2}{2n-2}L_{W}\left( \varrho _{+}\right) \right) \right\} .
\label{elecpot}
\end{eqnarray}%
The area law of the black hole entropy states that the entropy of a black
hole is the quarter of event horizon area \cite{Beck}. The entropy of almost
all kinds of black holes in Einstein gravity including dilaton black holes
is calculated by using this near universal law \cite{hunt}. Thus, the
entropy per unit volume $\omega _{n-1}$ of the Lifshitz black holes can be
found as 
\begin{equation}
s=\frac{r_{+}^{n-1}}{4}.  \label{entropy}
\end{equation}%
The Hawking temperature is 
\begin{equation*}
T=\frac{r_{+}^{z+1}f^{\prime }\left( r_{+}\right) }{4\pi l^{z+1}},
\end{equation*}%
which can be calculated as 
\begin{eqnarray}
T &=&\frac{(n+z-1)r_{+}^{z}}{4\pi l^{z+1}}+\allowbreak \frac{%
(n-2)^{2}kl^{1-z}}{4\pi (n+z-3)r_{+}^{2-z}}  \notag \\
&&-\frac{2\beta ^{2}l^{1-z}b^{2z-2}}{\pi (n-1)r_{+}^{z-2}}\left[ \frac{1}{2}+%
\frac{ql^{z-1}}{2\beta r_{+}^{n-1}}\left( \sqrt{L_{W}\left( \varrho
_{+}\right) }-\frac{1}{\sqrt{L_{W}\left( \varrho _{+}\right) }}\right) %
\right] .  \label{Temp}
\end{eqnarray}%
The large $\beta $ limit of temperature reproduces the temperature of
Einstein-Maxwell dilaton Lifshitz black holes \cite{kzsd}: 
\begin{eqnarray}
T &=&\frac{(n+z-1)r_{+}^{z}}{4\pi l^{z+1}}+\allowbreak \frac{%
(n-2)^{2}kl^{1-z}}{4\pi (n+z-3)r_{+}^{2-z}}-\frac{q^{2}l^{z-1}b^{2z-2}}{2\pi
(n-1)r_{+}^{2n+z-4}}+\frac{q^{4}l^{3z-3}b^{2z-2}}{8\pi
(n-1)r_{+}^{4n+z-6}\beta ^{2}}+O\left( \frac{1}{\beta ^{4}}\right) ,  \notag
\\
&&
\end{eqnarray}%
as one expects. We need the Smarr-type mass formula in order to check the
satisfaction of first law of thermodynamics. Using (\ref{mr+}), (\ref{Mass}%
), (\ref{charge}) and (\ref{entropy}), the Smarr-type mass can be written as 
\begin{eqnarray}
{M}\left( s,Q\right) &=&\frac{(n-1)\left( 4s\right) ^{(n+z-1)/(n-1)}}{16\pi
l^{z+1}}+\frac{\left( n-1\right) (n-2)^{2}k(4s)^{(n+z-3)/(n-1)}}{16\pi
l^{z-1}\left( n+z-3\right) ^{2}}  \notag \\
&&-\frac{\beta ^{2}b^{2z-2}(4s)^{(n-z+1)/(n-1)}\Pi }{2\pi l^{z-1}(n-z+1)},
\end{eqnarray}%
where 
\begin{eqnarray}
\Pi &=&\frac{1}{2}-\frac{(n-z+1)}{2(4s)^{(n-z+1)/(n-1)}}\left( \frac{16\pi
^{2}Q^{2}}{\beta ^{2}L_{W}\left( \zeta \right) }\right) ^{\left(
n-z+1\right) /\left( 2n-2\right) }  \notag \\
&&\times \left\{ \frac{L_{W}\left( \zeta \right) ^{2}}{3n+z-5}\mathbf{F}%
\left( \frac{3n+z-5}{2n-2},\frac{5n+z-7}{2n-2},\frac{z-2}{2n-2}L_{W}\left(
\zeta \right) \right) \right.  \notag \\
&&\left. +\frac{1}{n-z+1}\mathbf{F}\left( \frac{z-n-1}{2n-2},\frac{n+z-3}{%
2n-2},\frac{z-2}{2n-2}L_{W}\left( \zeta \right) \right) \right\} ,
\end{eqnarray}%
and $\zeta =\pi ^{2}Q^{2}/\left( \beta ^{2}s^{2}\right) $. As $\beta
\rightarrow \infty $, the behavior of Smarr-type mass is 
\begin{eqnarray}
M\left( s,Q\right) &=&\frac{(n-1)(4s)^{(n+z-1)/(n-1)}}{16\pi l^{z+1}}+\frac{%
(n-1)(n-2)^{2}k(4s)^{(n+z-3)/(n-1)}}{16\pi (n+z-3)^{2}l^{z-1}}  \notag \\
&&+\frac{2\pi Q^{2}b^{2z-2}(4s)^{(3-n-z)/(n-1)}}{(n+z-3)l^{z-1}}-\frac{8\pi
^{3}Q^{4}b^{2z-2}(4s)^{(5-3n-z)/(n-1)}}{(3n+z-5)l^{z-1}\beta ^{2}}+O\left( 
\frac{1}{\beta ^{2}}\right) .  \notag \\
&&
\end{eqnarray}%
This satisfy our expectation that the mass of the linear Maxwell case should
be recovered \cite{kzsd}. Now, in order to check the satisfaction of the
first law of thermodynamics, we take $S$ and $Q$\ as a complete set of
extensive quantities for mass $M(s,Q)$. Then, we define their conjugate
intensive quantities as temperature $T$\ and electric potential $U$ that
implies 
\begin{equation}
T=\left( \frac{\partial M}{\partial s}\right) _{Q}\text{ \ \ \ \ and \ \ \ \ 
}U=\left( \frac{\partial M}{\partial Q}\right) _{s}.  \label{intqua}
\end{equation}%
Intensive quantities computed by (\ref{intqua}) are in coincidance with ones
obtained by (\ref{elecpot}) and (\ref{Temp}). Hence, the first law of
thermodynamics 
\begin{equation}
dM=Tds+UdQ.
\end{equation}%
is satisfied. In the next subsection we investigate the stability of this
thermodynamic system under thermal perturbations.

\subsection{Thermal Stability in the Canonical and Grand-Canonical Ensembles}

\begin{figure*}[t]
\centering{%
\subfigure[~]{
   \label{fig2a}\includegraphics[width=.46\textwidth]{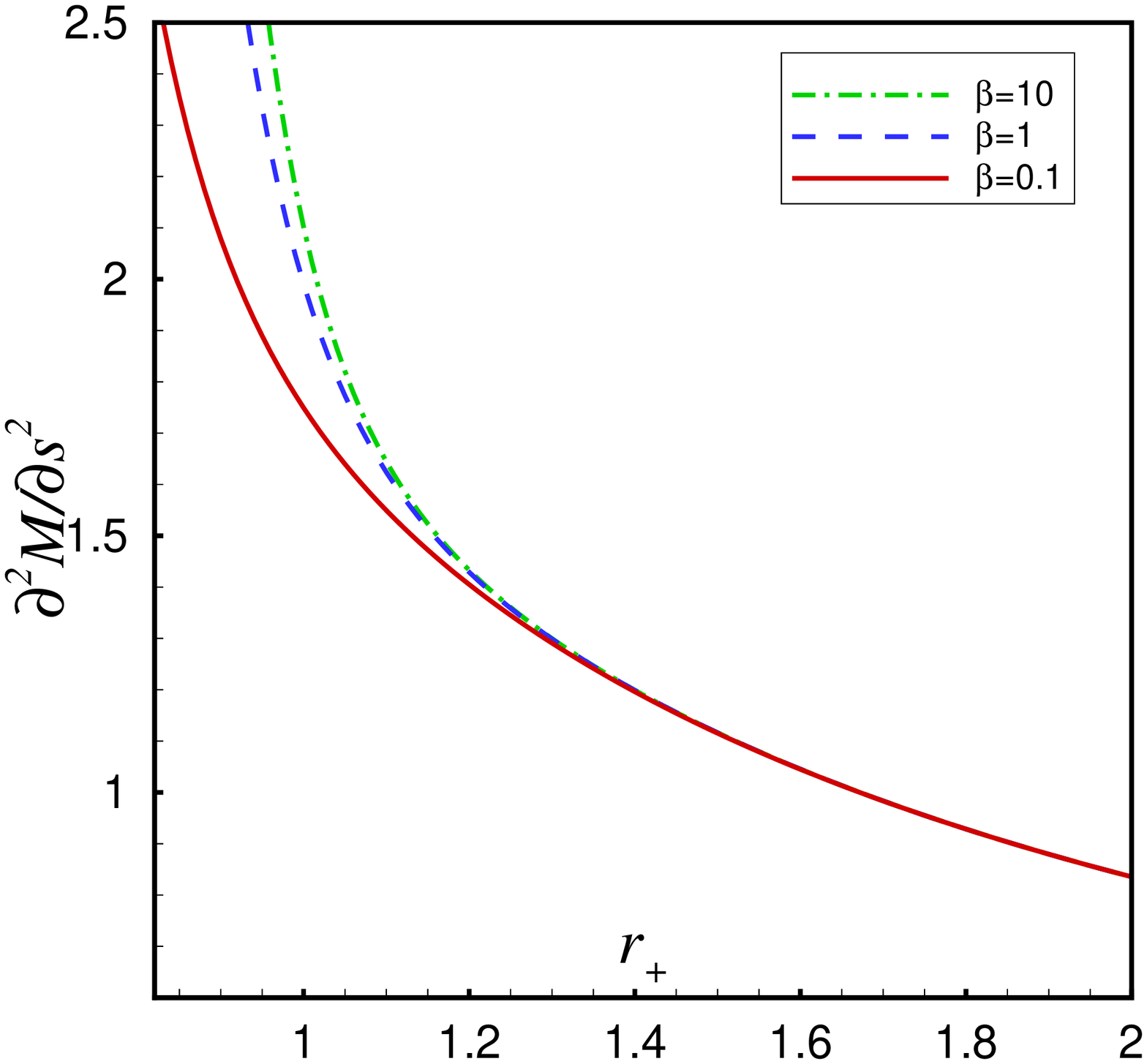}\qquad}} 
\subfigure[~]{
   \label{fig2b}\includegraphics[width=.46\textwidth]{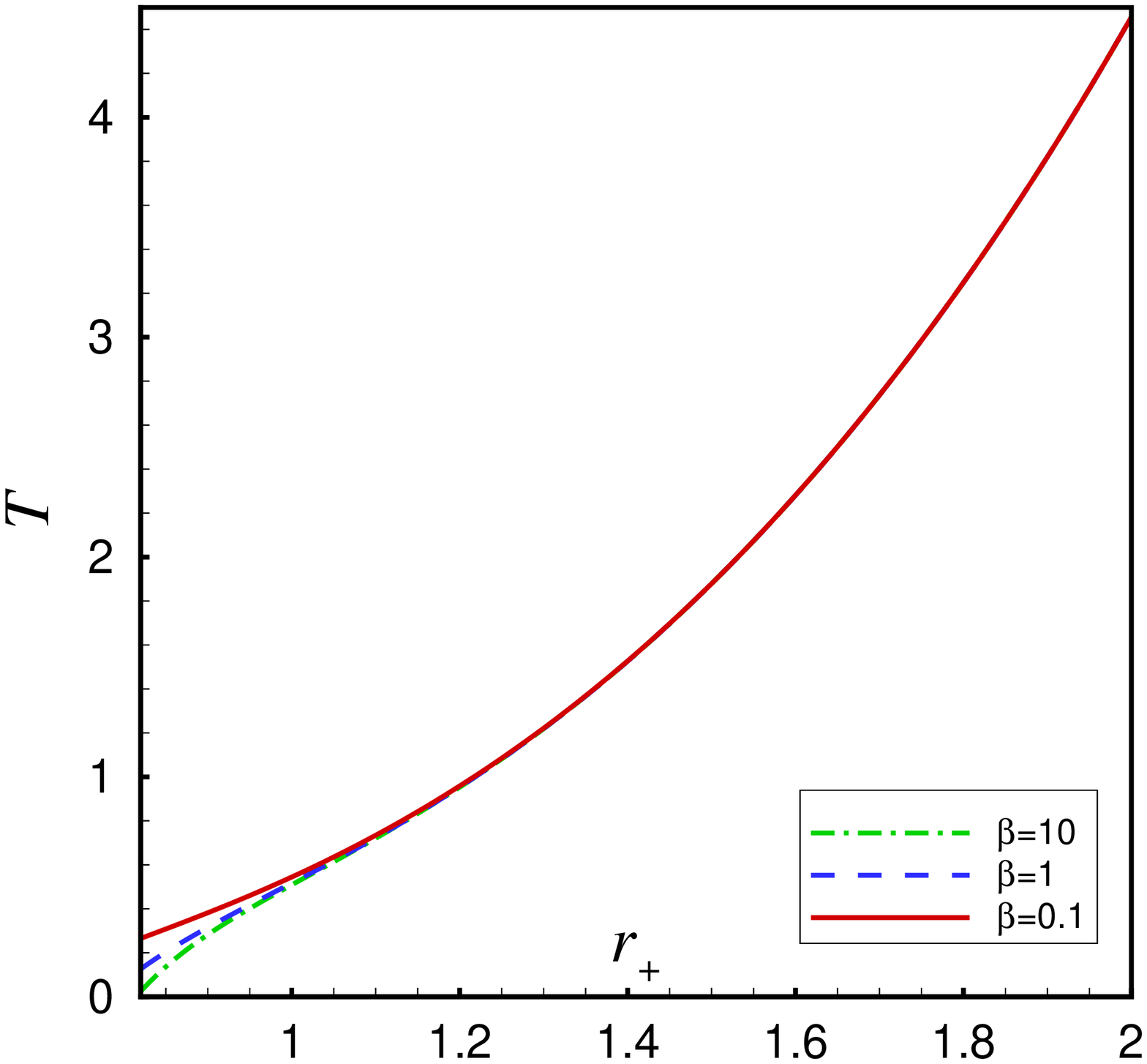}}
\caption{The behaviors of $\left( \partial ^{2}M/\partial s^{2}\right) _{Q}$
and $T$ versus $r_{+}$ for $k=0$ with $l=b=1$, $q=1.1$, $z=3$ and $n=5$.}
\label{fig2}
\end{figure*}
\begin{figure*}[t]
\centering{%
\subfigure[~]{
   \label{fig3a}\includegraphics[width=.46\textwidth]{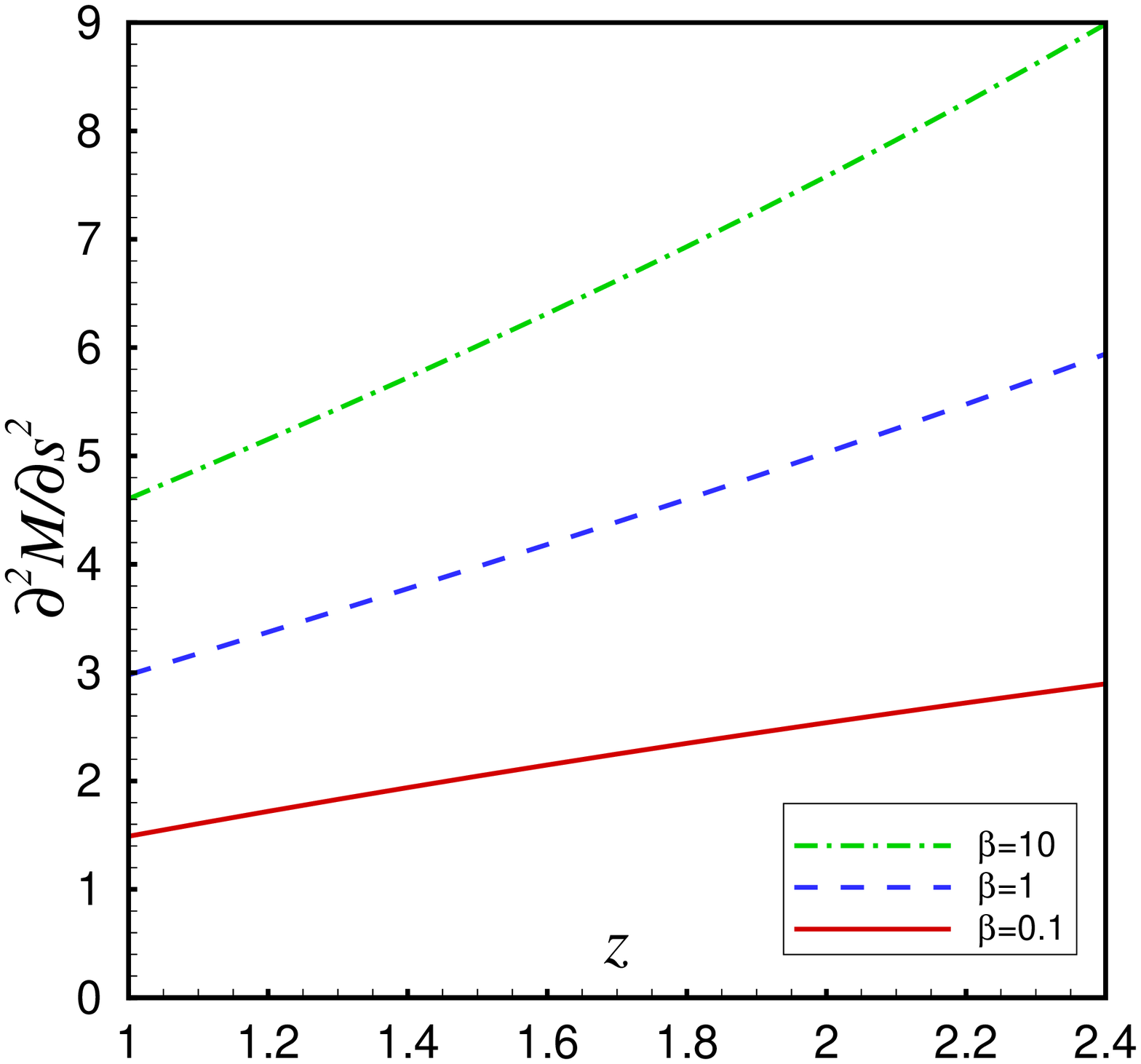}\qquad}} 
\subfigure[~]{
   \label{fig3b}\includegraphics[width=.46\textwidth]{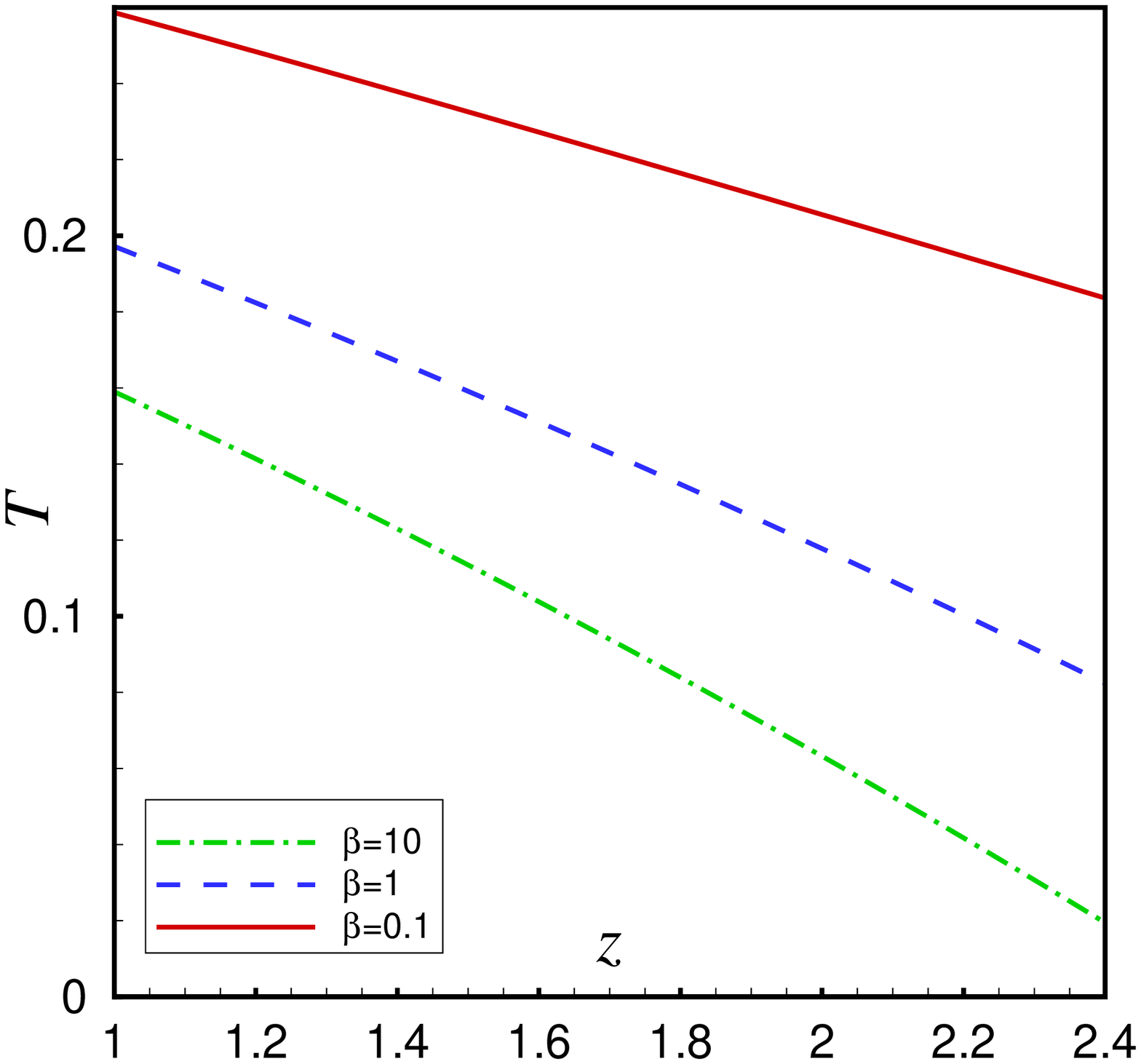}}
\caption{The behaviors of $\left( \partial ^{2}M/\partial s^{2}\right) _{Q}$
and $T$ versus $z$ for $k=0$ with $l=b=1$, $q=0.5$, $r_{+}=0.7$ and $n=5$.}
\label{fig3}
\end{figure*}
\begin{figure}[h]
\centerline{\includegraphics[width=.46\textwidth]{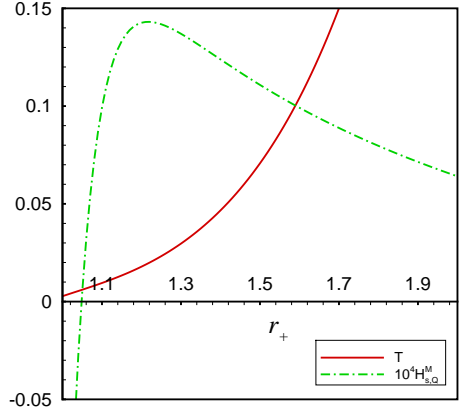}}
\caption{The behavior of $10^{4}\mathbf{H}_{s,Q}^{M}$ and $T$ versus $r_{+}$
for $k=0$ with $l=2$, $b=1$, $q=0.2$, $n=5$, $z=6$ and $\protect\beta =0.1$.}
\label{fig4}
\end{figure}
\begin{figure}[h]
\centerline{\includegraphics[width=.46\textwidth]{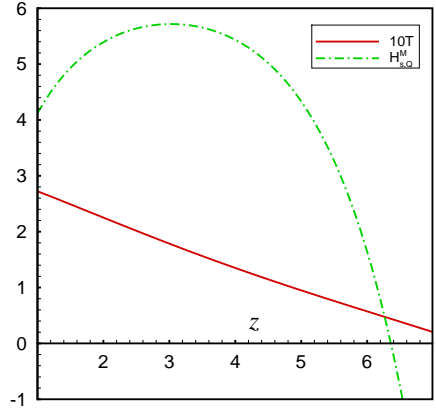}}
\caption{The behavior of $\mathbf{H}_{s,Q}^{M}$ and $10T$ versus $z$ for $%
k=0 $ with $l=b=1$, $q=0.2$, $n=5$, $r_{+}=0.7$ and $\protect\beta =0.1$.}
\label{fig5}
\end{figure}
\begin{figure}[h]
\centerline{\includegraphics[width=.46\textwidth]{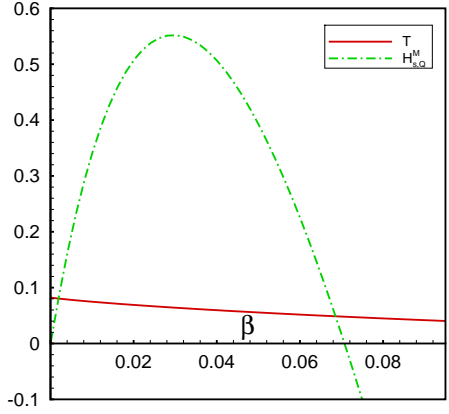}}
\caption{The behavior of $\mathbf{H}_{s,Q}^{M}$ and $T$ versus $\protect%
\beta $ for $k=0$ with $l=b=1$, $q=0.2$, $n=5$, $z=6.5$ and $r_{+}=0.7$.}
\label{fig6}
\end{figure}
\begin{figure}[h]
\centerline{\includegraphics[width=.46\textwidth]{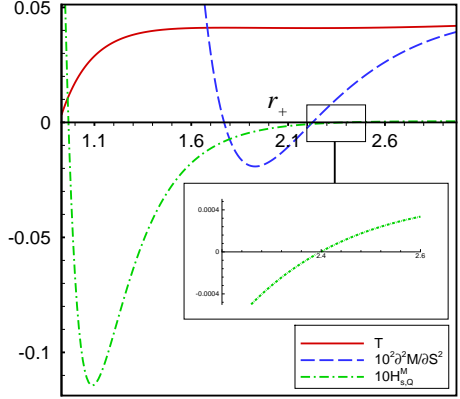}}
\caption{The behavior of $10^{2}(\partial ^{2}M/\partial s^{2})_{Q}$, $10%
\mathbf{H}_{s,Q}^{M}$ and $T$ versus $r_{+}$ for $k=1$ with $l=7$, $b=1$, $%
q=0.5$, $n=4$, $z=1.5$ and $\protect\beta =3$.}
\label{fig7}
\end{figure}
\begin{figure}[h]
\centerline{\includegraphics[width=.46\textwidth]{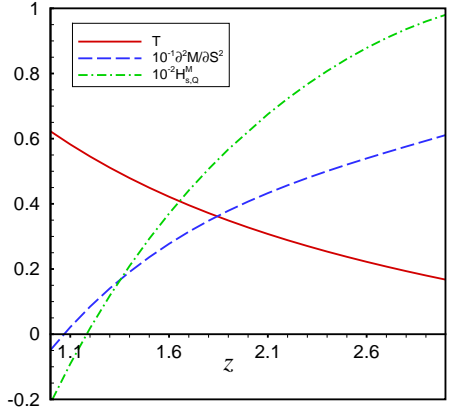}}
\caption{The behavior of $10^{-1}(\partial ^{2}M/\partial s^{2})_{Q}$, $%
10^{-2}\mathbf{H}_{s,Q}^{M}$ and $T$ versus $z$ for $k=1$ with $l=b=1$, $%
q=0.1$, $n=5$, $r_{+}=0.6$ and $\protect\beta =3$.}
\label{fig8}
\end{figure}
\begin{figure*}[t]
\centering{%
\subfigure[~]{
   \label{fig81a}\includegraphics[width=.46\textwidth]{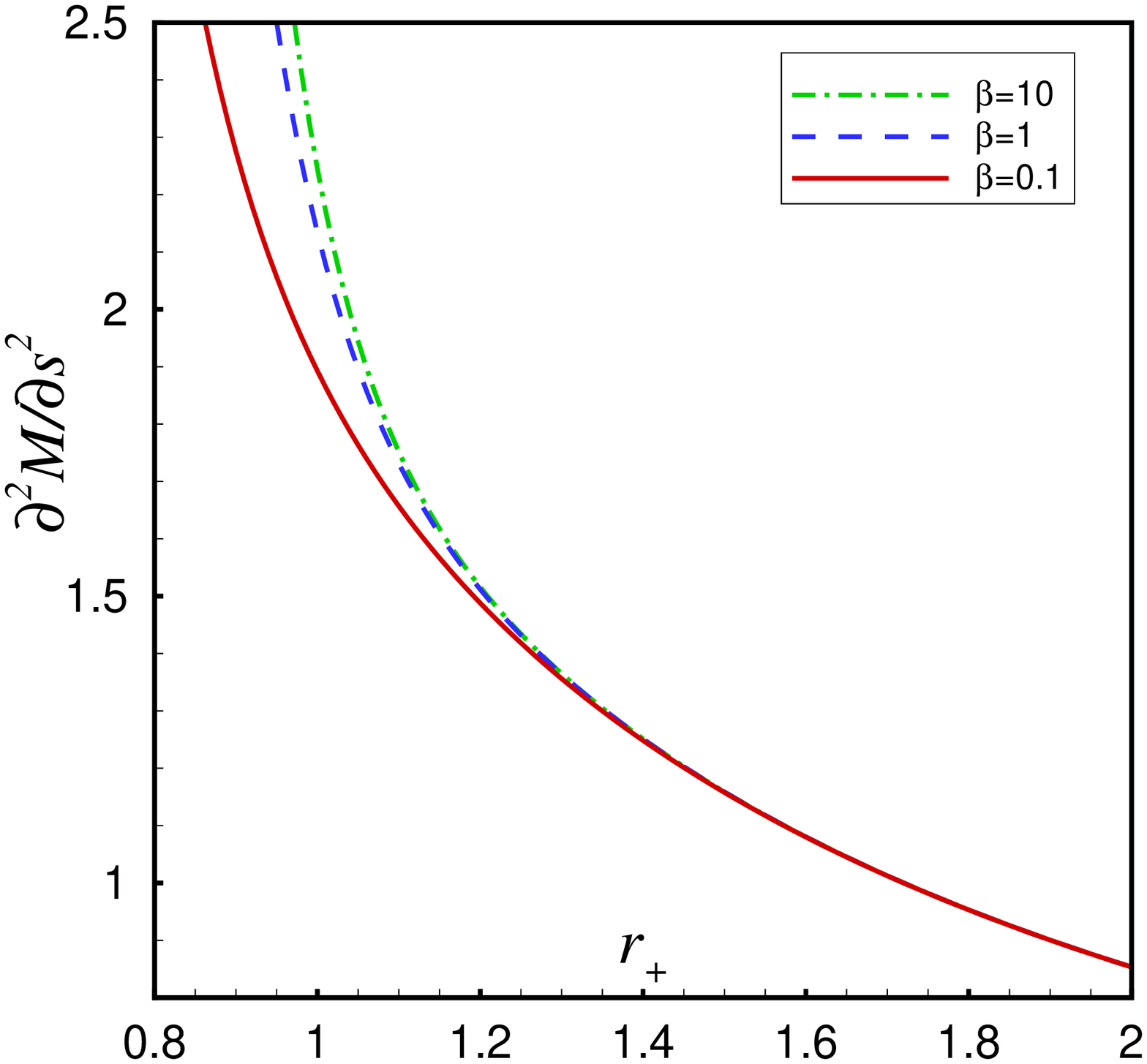}\qquad}} 
\subfigure[~]{
   \label{fig81b}\includegraphics[width=.46\textwidth]{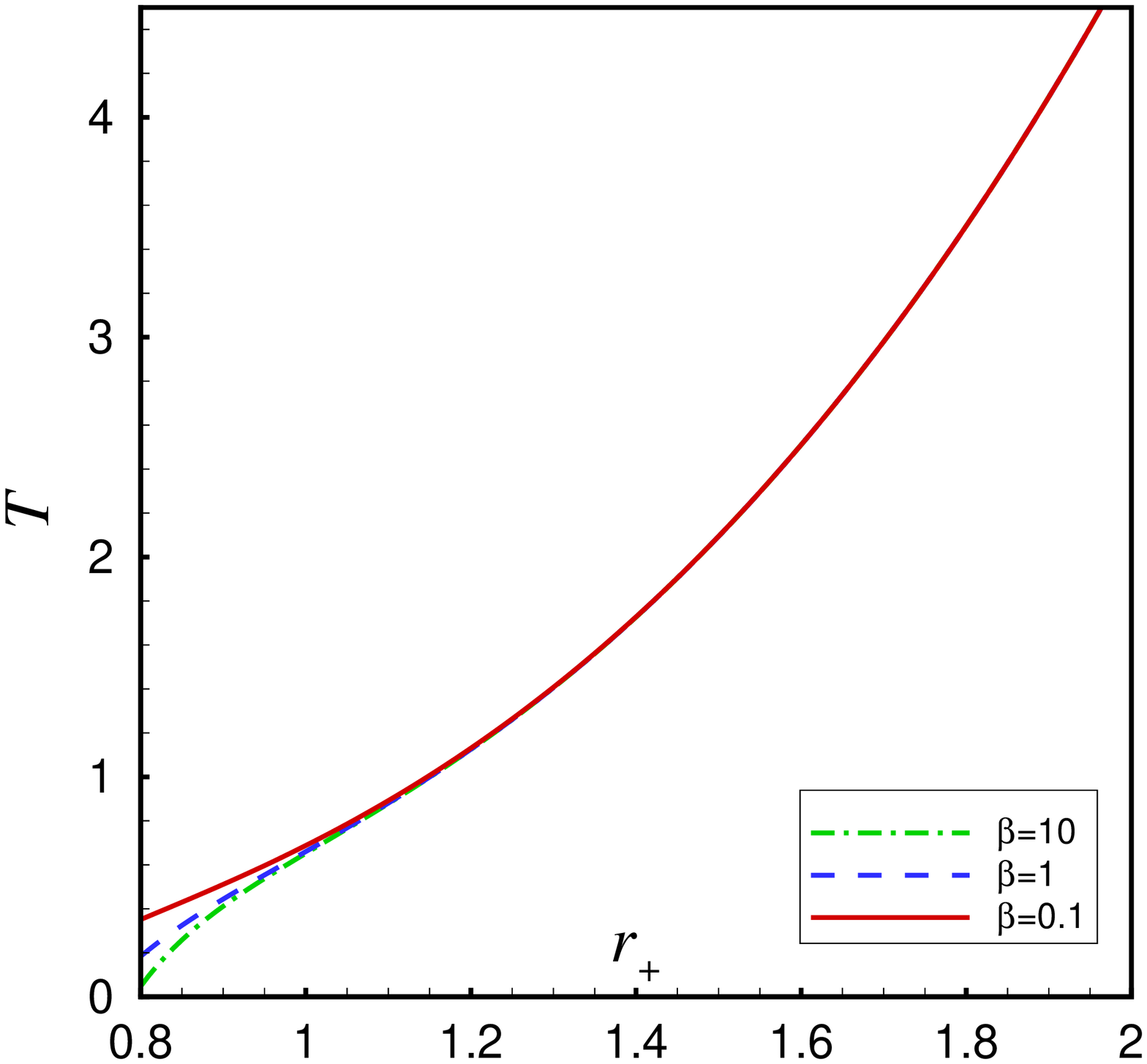}}
\caption{The behaviors of $\left( \partial ^{2}M/\partial s^{2}\right) _{Q}$
and $T$ versus $r_{+}$ for $k=1$ with $l=b=1$, $q=1.1$, $z=3$ and $n=5$.}
\label{fig8+1}
\end{figure*}
\begin{figure*}[t]
\centering{%
\subfigure[~]{
   \label{fig82a}\includegraphics[width=.46\textwidth]{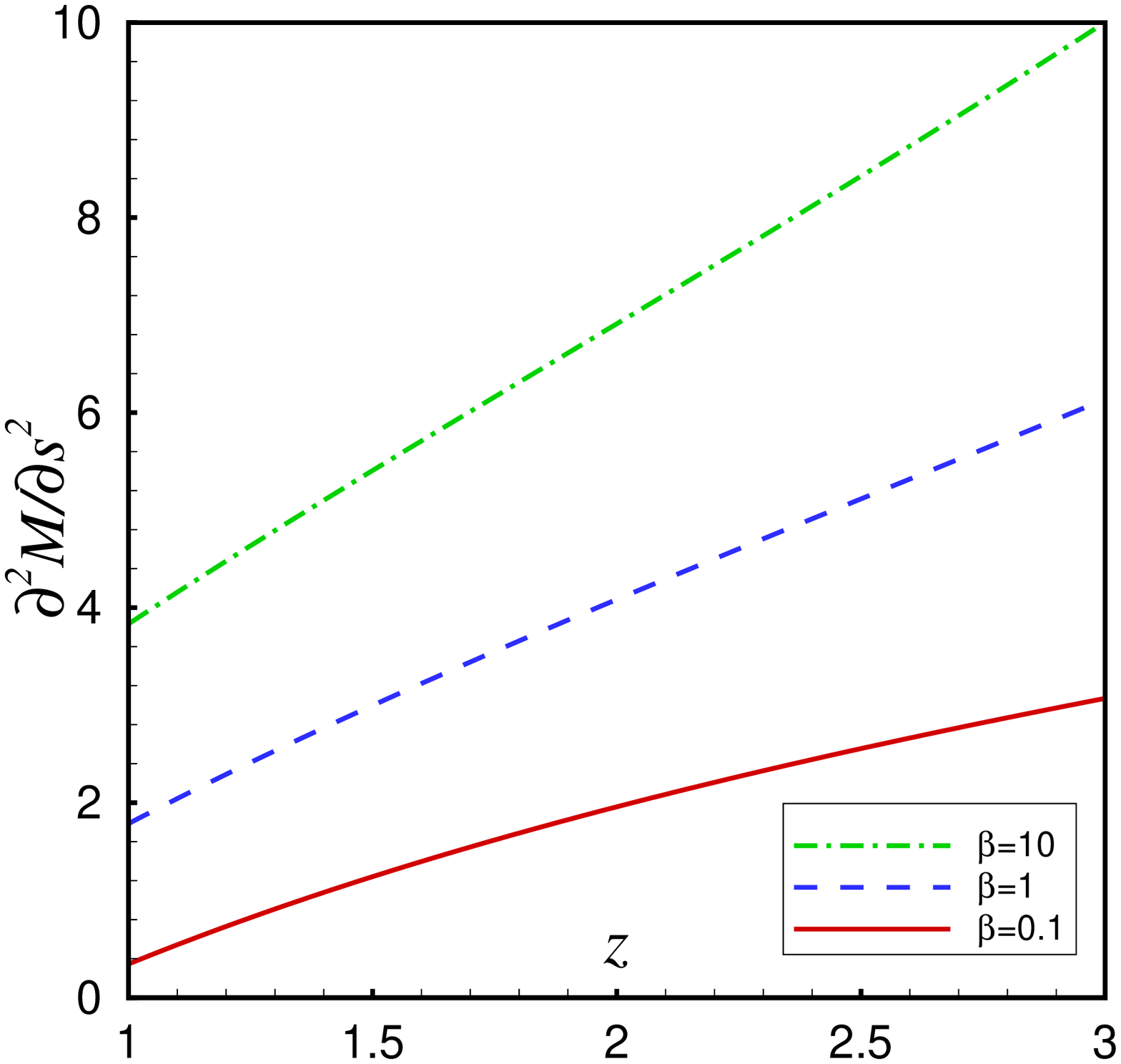}\qquad}} 
\subfigure[~]{
   \label{fig82b}\includegraphics[width=.46\textwidth]{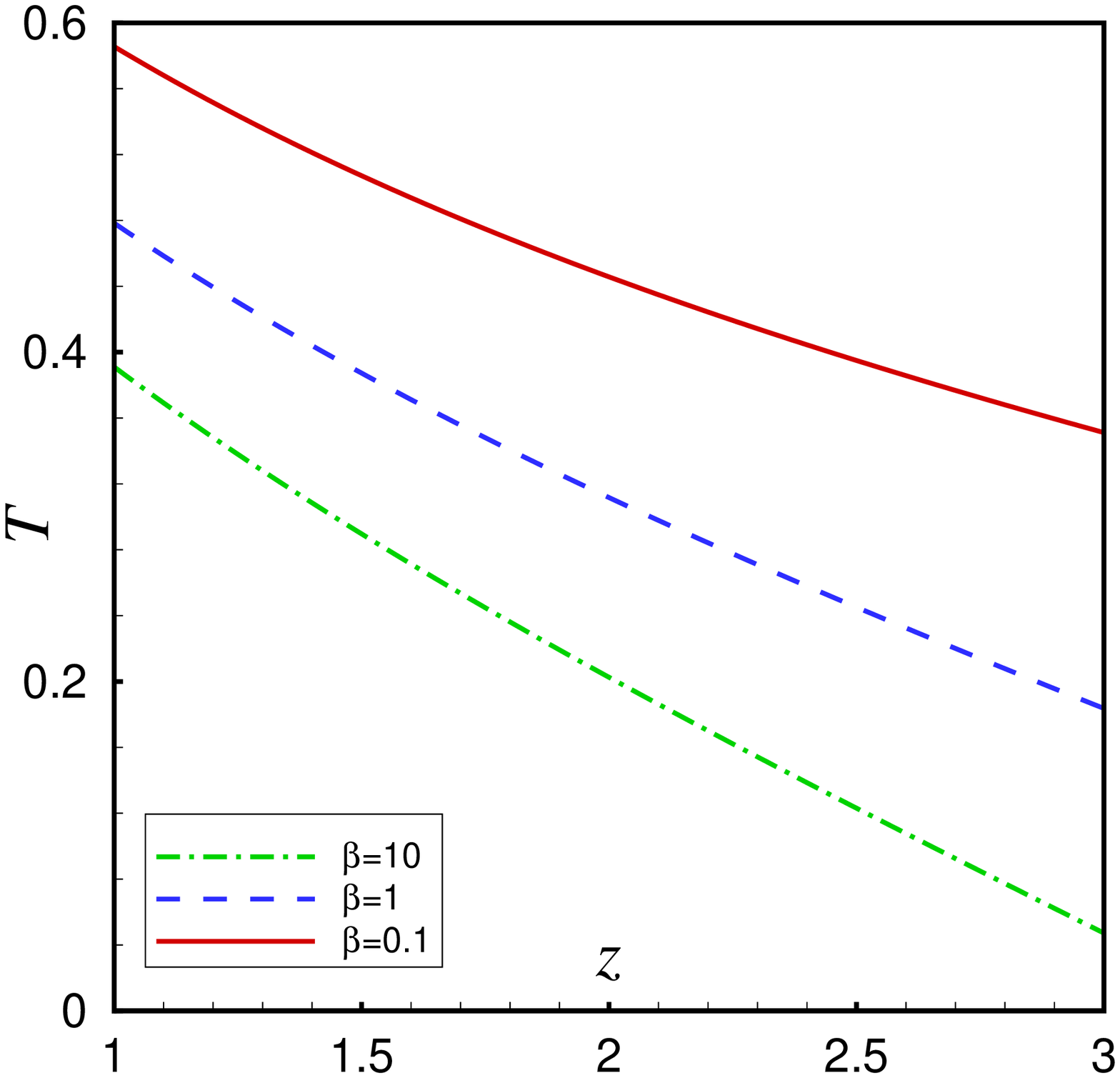}}
\caption{The behaviors of $\left( \partial ^{2}M/\partial s^{2}\right) _{Q}$
and $T$ versus $z$ for $k=1$ with $l=b=1$, $q=1.1$, $r_{+}=0.8$ and $n=5$.}
\label{fig8+2}
\end{figure*}
\begin{figure}[h]
\centerline{\includegraphics[width=.46\textwidth]{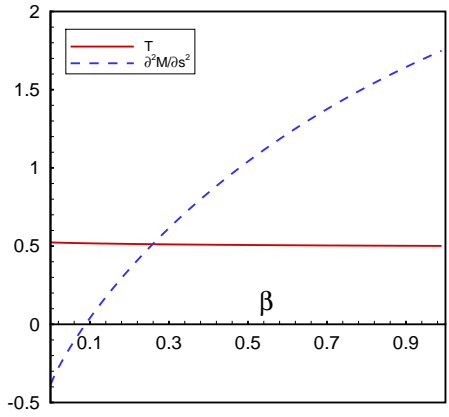}}
\caption{The behavior of $(\partial ^{2}M/\partial s^{2})_{Q}$ and $T$
versus $\protect\beta $ for $k=1$ with $l=1$, $b=0.3$, $q=0.1$, $n=5$, $%
r_{+}=0.5$ and $z=1.3$.}
\label{fig9}
\end{figure}
\begin{figure}[h]
\centerline{\includegraphics[width=.46\textwidth]{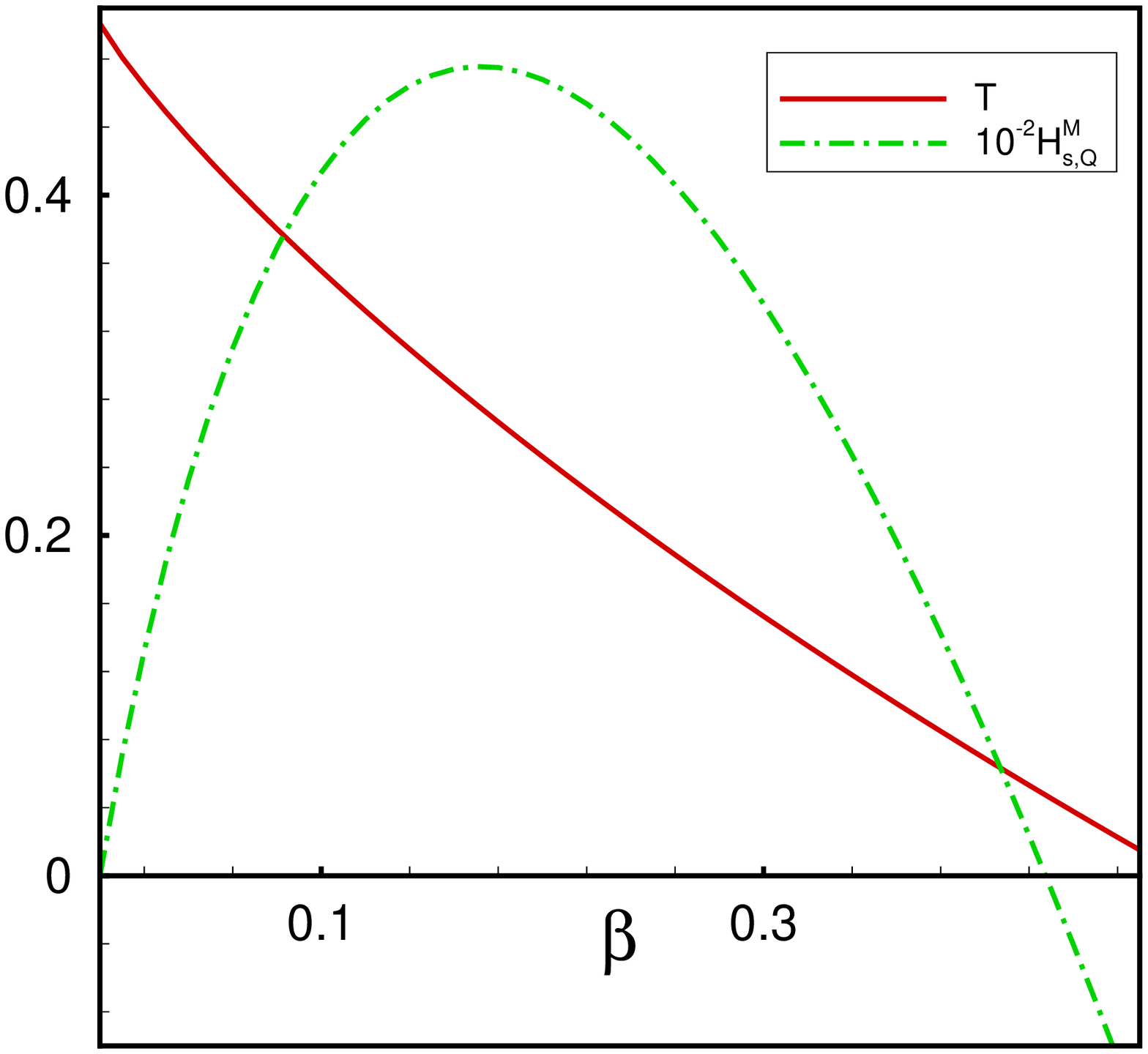}}
\caption{The behavior of $10^{-2}\mathbf{H}_{s,Q}^{M}$ and $T$ versus $%
\protect\beta $ for $k=1$ with $l=0.72$, $b=0.8$, $q=0.4$, $n=6$, $r_{+}=0.5$
and $z=3$.}
\label{fig10}
\end{figure}
In the first part of this section, we showed that one can regard a black
hole solution as a thermodynamic system. On the other hand, it is necessary
to investigate the stability of a thermodynamic system under thermal
perturbations. Therefore, this subsection is devoted to study the thermal
stability of the obtained solutions of the previous sections. The formal way
to analyze the stability of a thermodynamic system with respect to small
variations of the thermodynamic coordinates is by investigating the behavior
of the entropy $s(M,Q)$\ around equilibrium. This analysis can also be
performed in terms of the Legendre transformation of entropy namely $M(s,Q)$%
.\ In any ensemble, the local stability requires that the energy $M(s,Q)$\
be a convex function of its extensive variable \cite{Cal2,Gub}. The number
of thermodynamic variables is ensemble-dependent. In the canonical ensemble,
since the charge is fixed, the positivity of $(\partial ^{2}M/\partial
s^{2})_{Q}$\ in the ranges where temperature $T$ is positive suffices to
ensure the local stability. For the Lifshitz black solutions we have, 
\begin{eqnarray}
\left( \frac{\partial ^{2}M}{\partial s^{2}}\right) _{Q} &=&\frac{%
z(n-1+z)r_{+}^{z-n+1}}{(n-1)\pi l^{z+1}}+{\frac{k\left( n-2\right)
^{2}(z-2)r_{+}^{z-n-1}}{\pi l^{z-1}(n-1)\left( z+n-3\right) }}  \notag \\
&&+\frac{8(z-2)\beta ^{2}r_{+}^{3-n-z}}{\pi {(n-1)}^{2}l^{z-1}b^{2-2z}}%
\left\{ \frac{1}{2}+\frac{ql^{z-1}r_{+}^{1-n}}{2\beta (z-2)}\left[ (n+z-3)%
\sqrt{L_{W}\left( \varrho _{+}\right) }-\frac{z-2}{\sqrt{L_{W}\left( \varrho
_{+}\right) }}\right] \right\} .  \notag \\
&&  \label{dMSS}
\end{eqnarray}%
The behavior of (\ref{dMSS}) for large $\beta $ is 
\begin{eqnarray}
\left( \frac{\partial ^{2}M}{\partial s^{2}}\right) _{Q} &=&\frac{%
z(n-1+z)r_{+}^{z-n+1}}{(n-1)\pi l^{z+1}}+{\frac{k\left( n-2\right)
^{2}(z-2)r_{+}^{z-n-1}}{\pi l^{z-1}(n-1)\left( z+n-3\right) }}+\frac{%
2q^{2}(2n+z-4)b^{2z-2}r_{+}^{5-3n-z}}{\pi (n-1)^{2}l^{1-z}}  \notag \\
&&-\frac{q^{4}(4n+z-6)b^{2z-2}r_{+}^{7-5n-z}}{2\pi (n-1)^{2}l^{3-3z}\beta
^{2}}+O(\frac{1}{\beta ^{4}}),  \label{lbetadMdS}
\end{eqnarray}%
which recovers the result of \cite{kzsd} for the linear Maxwell
electrodynamics, as expected. Before we turn to grand canonical case it is
remarkable to find $\left( \partial ^{2}M/\partial s^{2}\right) _{Q}$for
highly nonlinear case i.e. $\beta \rightarrow 0$. In this case the behavior
of $\left( \partial ^{2}M/\partial s^{2}\right) _{Q}$is 
\begin{eqnarray}
\lim_{\beta \rightarrow 0}\left( \frac{\partial ^{2}M}{\partial s^{2}}%
\right) _{Q} &=&\frac{z(n-1+z)r_{+}^{z-n+1}}{(n-1)\pi l^{z+1}}+{\frac{%
k\left( n-2\right) ^{2}(z-2)r_{+}^{z-n-1}}{\pi l^{z-1}(n-1)\left(
z+n-3\right) }}  \notag \\
&&+\frac{4q\beta r_{+}^{4-2n-z}(n+z-3)}{\pi {(n-1)}^{2}b^{2-2z}}\sqrt{%
L_{W}\left( \frac{q^{2}l^{2z-2}}{\beta ^{2}r_{+}^{2n-2}}\right) }.
\label{sbetadMdS}
\end{eqnarray}%
Soon we use (\ref{sbetadMdS}) to discuss thermal stability of highly
nonlinear solutions. In the grand-canonical ensemble, since $Q$\ is not a
fixed parameter, the local stability requires the positivity of determinant
of Hessian matrix $\mathbf{H}_{sQ}^{M}=\left[ \partial ^{2}M/\partial
s\partial Q\right] >0$. The determinant of the Hessian matrix $\mathbf{H}%
_{s,Q}^{M}$ for the solutions under consideration can be calculated as 
\begin{eqnarray}
\mathbf{H}_{s,Q}^{M} &=&\frac{4\beta b^{2z-2}l^{3-3z}}{q(n-1)^{2}r_{+}^{n-3}}%
\sqrt{L_{W}\left( \eta _{+}\right) }\left[ \Upsilon +\frac{4q\beta
(z-2)b^{2z-2}l^{z-1}}{(n-1)r_{+}^{n+2z-3}}\left( \sqrt{L_{W}\left( \eta
_{+}\right) }-\frac{1}{\sqrt{L_{W}\left( \eta _{+}\right) }}\right) \right] 
\notag \\
&&-\frac{16b^{2z-2}l^{2-2z}(z-2)}{r_{+}^{n-z-1}(2n-2)^{2}}\left( \frac{\beta
^{2}L_{W}\left( \eta _{+}\right) }{q^{2}l^{2z-2}}\right) ^{(n+z-3)/(2n-2)} 
\notag \\
&&\times \left[ \Upsilon +\frac{4q\beta b^{2z-2}l^{z-1}}{(n-1)r_{+}^{n+2z-3}}%
\left( (n+z-3)\sqrt{L_{W}\left( \eta _{+}\right) }-\frac{z-2}{\sqrt{%
L_{W}\left( \eta _{+}\right) }}\right) \right]  \notag \\
&&\times \left\{ \frac{L_{W}\left( \eta _{+}\right) }{3n+z-5}\mathbf{F}%
\left( \frac{3n+z-5}{2n-2},\frac{5n+z-7}{2n-2},\frac{z-2}{2n-2}L_{W}\left(
\zeta \right) \right) \right.  \notag \\
&&\left. +\frac{1}{n+z-3}\mathbf{F}\left( \frac{n+z-3}{2n-2},\frac{3n+z-5}{%
2n-2},\frac{z-2}{2n-2}L_{W}\left( \zeta \right) \right) \right\} ,
\end{eqnarray}%
where%
\begin{equation*}
\Upsilon =\frac{z(n+z-1)}{l^{2}}+\frac{(z-2)(n-2)^{2}k}{(n+z-3)r_{+}^{2}}+%
\frac{4(z-2)\beta ^{2}b^{2z-2}}{(n-1)r_{+}^{2z-2}}.
\end{equation*}%
The large $\beta $ behavior of $\mathbf{H}_{sQ}^{M}$ is 
\begin{eqnarray}
\mathbf{H}_{s,Q}^{M} &=&\frac{4z(n+z-1)b^{2z-2}r_{+}^{4-2n}}{%
(n-1)(n+z-3)l^{2z}}+\frac{4(n-2)^{2}(z-2)kb^{2z-2}r_{+}^{2-2n}}{%
(n-1)(n+z-3)^{2}l^{2z-2}}  \notag \\
&&-\frac{8q^{2}(z-2)b^{4z-4}r_{+}^{8-4n-2z}}{(n-1)^{2}(n+z-3)}+\frac{%
2q^{2}b^{2z-2}}{(n-1)(3n+z-5)\beta ^{2}r_{+}^{4n-4}}  \notag \\
&&\times \left( \frac{q^{2}(z-2)b^{2z-2}l^{2z-2}}{(n-1)(n+z-3)r_{+}^{2n+2z-6}%
}-\frac{3k(n-2)^{3}}{n+z-3}-\frac{3z(n+z-1)r_{+}^{2}}{l^{2}}\right) +O\left( 
\frac{1}{\beta ^{4}}\right) ,  \notag \\
&&
\end{eqnarray}%
which coincides with the one of linear Maxwell case \cite{kzsd}. In what
follows we discuss the stability for $k=0$ (black branes) and $k=1$
(spherical black holes), separately.

$k=0$: In this case, calculations show that the solutions are always stable
in the canonical ensemble. Although it is difficult to see this for
arbitrary $\beta $ from (\ref{dMSS}), we can show this fact for large and
small $\beta $'s. For large $\beta $ where the linear Maxwell regime is
dominant, this fact that the solutions are always stable in canonical
ensemble has been pointed out in \cite{kzsd}. For small $\beta $, the
behaviour of $\left( \partial ^{2}M/\partial s^{2}\right) _{Q}$ has been
given in (\ref{sbetadMdS}). It is obvious from (\ref{sbetadMdS}) that $%
\lim_{\beta \rightarrow 0}\left( \partial ^{2}M/\partial s^{2}\right) _{Q}>0$
for $k=0$. Figs. \ref{fig2a} and \ref{fig3a} depict the fact that black
branes are always stable in canonical ensemble. Of course, one should check
the positivity of temperature for these choices. This can be seen in Figs. %
\ref{fig2b} and \ref{fig3b} which show that the temperature is positive for
these choices and therefore we have black branes. In grand canonical
ensemble, the system is thermally stable provided that the radius of the
black hole is larger than $r_{+\min }$ (Fig. \ref{fig4}). Figure \ref{fig5}
shows that there is a $z_{\max }$ that for values greater than it we
encounter instability. We also have a $\beta _{\max }$ that black branes are
stable under thermal perturbations for values lower than it as one can see
in Fig. \ref{fig6}.

$k=1$: As one can see from Fig. \ref{fig7}, there is a Hawking-Page phase
transition between small and large black holes in both ensembles. In terms
of $z$ we have an ensemble-dependent minimum $z_{\min }$ which depends on
the parameters of the system and black holes are unstable for values lower
than it as Fig. \ref{fig8} shows. The value of $z_{\min }$ is greater in
grand-canonical ensemble compare to that in canonical ensemble. For highly
nonlinear case (small $\beta $) and also linear case (large $\beta $) black
holes are stable for $z\geq 2$ in canonical ensemble as one can see from (%
\ref{lbetadMdS}) and (\ref{sbetadMdS}). Hence, one can conclude that there
are always stable black holes for $z\geq 2$ in canonical ensemble (Figs. \ref%
{fig8+1} and \ref{fig8+2}). Fig. \ref{fig9} shows that black holes are
unstable in canonical ensemble for $\beta <\beta _{\min }$ in the case of $%
z<2$. In grand-canonical ensemble we have stable black holes for $\beta
<\beta _{\max }$ (Fig. \ref{fig10}).

\section{Gauge/Gravity Duality\label{GGD}}

In this section, we would like to perform the gauge/gravity duality idea on
our obtained solutions. We first try to calculate zero-frequency shear
viscosity for a hydrodynamic system. Then, we employ this idea to study the
behavior of the holographic conductivity for a two dimensional system for
both asymptotic AdS ($z=1$) and asymptotic Lifshitz cases and present
experimental observations matched with obtained results.

\subsection{Holographic Viscosity}

Here we intend to calculate the ratio of shear viscosity to entropy $\eta /s$
in the zero frequency limit. For this aim, we use the pole method \cite%
{vis1,vis2}. We consider the five-dimensional planar metric%
\begin{equation}
ds^{2}=-\frac{r^{2z}f(r)}{l^{2z}}dt^{2}+{\frac{l^{2}dr^{2}}{r^{2}f(r)}}%
+r^{2}(dx_{1}^{2}+dx_{2}^{2}+dx_{3}^{2}),  \label{plmet}
\end{equation}%
where the QFT lives on the $4$-dimensional $r$-infinity boundary. Defining $%
y=1-r_{+}^{2}/r^{2}$, one can rewrite (\ref{plmet}) as%
\begin{equation}
ds^{2}=-\frac{F(y)r_{+}^{2z}}{l^{2z}(1-y)^{z}}dt^{2}+{\frac{l^{2}dy^{2}}{%
4F(y)(1-y)^{2}}}+\frac{r_{+}^{2}}{1-y}(dx_{1}^{2}+dx_{2}^{2}+dx_{3}^{2}),
\label{plmet2}
\end{equation}%
where

\begin{eqnarray*}
F(y) &=&1-\frac{m(1-y)^{(3+z)/2}}{r_{+}^{3+z}}-\frac{4\beta
^{2}l^{2}b^{2z-2}(1-y)^{z-1}}{3(5-z)r_{+}^{2z-2}}+\frac{4\beta
^{2}l^{2}b^{2z-2}(1-y)^{(3+z)/2}}{3r_{+}^{3+z}}\left( \frac{q^{2}l^{2z-2}}{%
\beta ^{2}L_{W}(\varrho _{y})}\right) ^{\left( 5-z\right) /6} \\
&&\times \left\{ \frac{L_{W}^{2}(\varrho _{y})}{7+z}\mathbf{F}\left( \frac{%
7+z}{6},\frac{13+z}{6},\frac{z-2}{6}L_{W}(\varrho _{y})\right) +\frac{1}{5-z}%
\mathbf{F}\left( \frac{z-5}{6},\frac{z+1}{6},\frac{z-2}{6}L_{W}(\varrho
_{y})\right) \right\} ,
\end{eqnarray*}%
$\varrho _{y}\equiv q^{2}l^{2z-2}(1-y)^{3}/(\beta ^{2}r_{+}^{6})$ and
horizon located at $y=0$. In order to use the pole method, we first apply an
off-shell perturbation%
\begin{equation}
dx_{i}\rightarrow dx_{i}+\varepsilon e^{-i\omega t}dx_{j},
\end{equation}%
where $\varepsilon $ is an infinitesimal positive parameter. Then, by using
the perturbed metric and the expansion of $F(y)$ near the horizon, namely 
\begin{eqnarray}
F(y) &=&\left[ \frac{z+3}{2}-\frac{2\beta ^{2}b^{2z-2}l^{2}}{3r_{+}^{2z-2}}-%
\frac{2q\beta b^{2z-2}l^{z+1}}{3r_{+}^{2z+1}}\left( \sqrt{L_{W}\left(
\varrho _{+}\right) }-\frac{1}{\sqrt{L_{W}\left( \varrho _{+}\right) }}%
\right) \right] y \\
&&-\left[ \frac{(z+3)(z+1)}{8}-\frac{(3z+2)q^{2}b^{2z-2}l^{2z}}{6r_{+}^{2z+4}%
}\exp \left[ -\frac{1}{2}L_{W}\left( \varrho _{+}\right) \right] \right. 
\notag \\
&&\left. +\frac{(3z-1)\beta ^{2}b^{2z-2}l^{2}}{6r_{+}^{2z-2}}\left( \exp %
\left[ -\frac{1}{2}L_{W}\left( \varrho _{+}\right) \right] -1\right) \right]
y^{2}+O\left( y^{3}\right) ,
\end{eqnarray}%
we calculate the residue of the pole at $y=0$ in the Lagrangian density%
\begin{equation}
\text{\textrm{Res}}_{y=0}\mathcal{L}=-\frac{\varepsilon ^{2}\omega
^{2}r_{+}^{3}e^{-2i\omega t}}{128\pi ^{2}T},  \label{ResL}
\end{equation}%
where $T$ is given by Eq. (\ref{Temp}). Now, via the formula \cite{vis1,vis2}%
\begin{equation}
\eta =-8\pi T\lim_{\varepsilon ,\omega \rightarrow 0}\frac{\text{\textrm{Res}%
}_{y=0}\mathcal{L}}{\varepsilon ^{2}\omega ^{2}},
\end{equation}%
the shear viscosity for zero frequency case can be obtained as%
\begin{equation}
\eta =\frac{r_{+}^{3}}{16\pi }.
\end{equation}%
Therefore, the ratio $\eta /s$ is obtained as 
\begin{equation}
\frac{\eta }{s}=\frac{1}{4\pi },  \label{shvis}
\end{equation}%
where we have used Eq. (\ref{entropy}). Therefore, we regain the well-known
value $1/4\pi $ for $\eta /s$ as many cases in Einstein gravity. This
implies that neither non-AdS symmetry of the system on boundary nor the
presence of the dilaton and additional gauge fields cannot affect the value
of the shear viscosity of the system.

\subsection{Holographic Conductivity}

Our aim in this subsection is to calculate the holographic conductivity by
performing gauge/gravity duality for both linear Maxwell (infinite $\beta $)
and nonlinear cases. Finally, we depict the behaviour of conductivity for
linear and nonlinear electrodynamics cases and show that our results are
supported by some experimental observations.

\subsubsection{Linear Maxwell case ($\protect\beta$-infinity)}

We consider a four-dimensional planar metric%
\begin{equation}
ds^{2}=-\frac{\mathcal{F}_{\infty }(u)r_{+}^{4z}}{l^{2z}u^{2z}}dt^{2}+{\frac{%
l^{2}du^{2}}{\mathcal{F}_{\infty }(u)u^{2}}}+\frac{r_{+}^{4}}{u^{2}}%
(dx_{1}^{2}+dx_{2}^{2}),  \label{plmet3}
\end{equation}%
where

\begin{equation}
\mathcal{F}_{\infty }(u)=1-\frac{mu^{z+2}}{r_{+}^{2z-4}}+\frac{%
q^{2}b^{2z-2}l^{2z}u^{2z+2}}{zr_{+}^{4z+4}},  \notag
\end{equation}%
which can be obtained by defining $u=r_{+}^{2}/r$ in the metric (\ref{met}).
The horizon locates at $u=r_{+}$, while the QFT lives at $u=0$ boundary. We
turn on perturbations $g_{tx_{1}}\left( u\right) e^{-i\omega t}$ and $%
A_{x_{1}}\left( u\right) e^{-i\omega t}$ and therefore receive one Einstein
and one electrodynamic additional equations of motion as {\small 
\begin{equation}
A_{x_{1}}^{\prime \prime }\left( u\right) +\left[ \frac{\mathcal{F}_{\infty
}^{\prime }(u)}{\mathcal{F}_{\infty }(u)}+\frac{3(1-z)}{u}\right]
A_{x_{1}}^{\prime }\left( u\right) +\frac{l^{2z+2}u^{2z-2}}{r_{+}^{4z}%
\mathcal{F}_{\infty }(u)^{2}}\left[ \omega ^{2}A_{x_{1}}\left( u\right) -%
\frac{qb^{2z-2}u^{z}\mathcal{F}_{\infty }(u)}{l^{2}r_{+}^{2z}}\left(
2g_{tx_{1}}\left( u\right) +ug_{tx_{1}}^{\prime }\left( u\right) \right) %
\right] =0,  \label{E10}
\end{equation}%
} and%
\begin{equation}
2g_{tx_{1}}\left( u\right) +ug_{tx_{1}}^{\prime }\left( u\right)
=4qr_{+}^{2z-4}u^{2-z}A_{x_{1}}\left( u\right) ,  \label{E2}
\end{equation}%
where the prime denotes derivative with respect to $u$. Combining Eqs. (\ref%
{E10}) and (\ref{E2}), one can easily find a decoupled equation for $%
A_{x_{1}}$

\begin{equation}
A_{x_{1}}^{\prime \prime }\left( u\right) +\left[ \frac{\mathcal{F}_{\infty
}^{\prime }(u)}{\mathcal{F}_{\infty }(u)}+\frac{3(1-z)}{u}\right]
A_{x_{1}}^{\prime }\left( u\right) +\frac{l^{2z+2}u^{2z-2}}{r_{+}^{4z}%
\mathcal{F}_{\infty }(u)^{2}}\left[ \omega ^{2}-\frac{4q^{2}b^{2z-2}u^{2}%
\mathcal{F}_{\infty }(u)}{l^{2}r_{+}^{4}}\right] A_{x_{1}}\left( u\right) =0.
\label{E3}
\end{equation}%
Near the boundary $u=0$, Eq. (\ref{E3}) reduces approximately to 
\begin{equation}
A_{x_{1}}^{\prime \prime }\left( u\right) +\frac{3(1-z)}{u}A_{x_{1}}^{\prime
}\left( u\right) +\cdots =0,  \label{nearbound}
\end{equation}%
which has the solution%
\begin{equation}
A_{x_{1}}\left( u\right) =A^{0}+A^{1}u^{3z-2}+\cdots ,  \label{nearboundsol}
\end{equation}%
where $A^{0}$ and $A^{1}$ are constants of integration. Following the
procedure presented in appendix \ref{app1}, we can calculate the
conductivity as

\begin{equation}
\sigma =\frac{(3z-2)r_{+}^{6z-4}A^{1}}{4\pi i\omega b^{2z-2}l^{z+1}A^{0}}
\end{equation}

\subsubsection{Nonlinear electrodynamics}

Here we follow the above procedure to calculate the holographic conductivity
in the case of nonlinear electrodynamics. We consider the four-dimensional
planar metric%
\begin{equation}
ds^{2}=-\frac{\mathcal{F}(u)r_{+}^{4z}}{l^{2z}u^{2z}}dt^{2}+{\frac{%
l^{2}du^{2}}{\mathcal{F}(u)u^{2}}}+\frac{r_{+}^{4}}{u^{2}}%
(dx_{1}^{2}+dx_{2}^{2}),
\end{equation}%
where

\begin{eqnarray}
\mathcal{F}(u) &=&1-\frac{mu^{z+2}}{r_{+}^{2z+4}}-\frac{2\beta
^{2}l^{2}b^{2z-2}u^{2z-2}}{(4-z)r_{+}^{4z-4}}+\frac{2\beta
^{2}l^{2}b^{2z-2}u^{z+2}}{r_{+}^{2z+4}}\left( \frac{q^{2}l^{2z-2}}{\beta
^{2}L_{W}(\varrho _{u})}\right) ^{\left( 4-z\right) /4}  \notag \\
&&\times \left\{ \frac{L_{W}^{2}(\varrho _{u})}{4+z}\mathbf{F}\left( \frac{%
4+z}{4},\frac{8+z}{4},\frac{z-2}{4}L_{W}(\varrho _{u})\right) +\frac{1}{4-z}%
\mathbf{F}\left( \frac{z-4}{4},\frac{z}{4},\frac{z-2}{4}L_{W}(\varrho
_{u})\right) \right\} ,  \notag
\end{eqnarray}%
and $\varrho _{u}\equiv q^{2}l^{2z-2}u^{4}/r_{+}^{8}\beta ^{2}$. Turning on
perturbations $g_{tx_{1}}\left( u\right) e^{-i\omega t}$ and $%
A_{x_{1}}\left( u\right) e^{-i\omega t}$, we find below equations of motion

\begin{gather}
A_{x_{1}}^{\prime \prime }\left( u\right) +\left[ \frac{\mathcal{F}^{\prime
}(u)}{\mathcal{F}(u)}+\frac{3(1-z)}{u}+\frac{2L_{W}\left( \varrho
_{u}\right) }{u(1+L_{W}\left( \varrho _{u}\right) )}\right]
A_{x_{1}}^{\prime }\left( u\right)  \notag \\
+\frac{l^{2z+2}u^{2z-2}}{r_{+}^{4z}\mathcal{F}(u)^{2}}\left[ \omega
^{2}A_{x_{1}}\left( u\right) -\frac{\beta b^{2z-2}u^{z-2}\mathcal{F}(u)}{%
l^{z+1}r_{+}^{2z-4}}\sqrt{L_{W}\left( \varrho _{u}\right) }\left(
2g_{tx_{1}}\left( u\right) +ug_{tx_{1}}^{\prime }\left( u\right) \right) %
\right] =0,  \label{E1}
\end{gather}%
and%
\begin{equation}
2g_{tx_{1}}\left( u\right) +ug_{tx_{1}}^{\prime }\left( u\right)
=4qr_{+}^{2z-4}u^{2-z}A_{x_{1}}\left( u\right) ,  \label{E21}
\end{equation}%
The result of combining Eqs. (\ref{E1}) and (\ref{E21}) is decoupled
equation for $A_{x_{1}}$%
\begin{gather}
A_{x_{1}}^{\prime \prime }\left( u\right) +\left[ \frac{\mathcal{F}^{\prime
}(u)}{\mathcal{F}(u)}+\frac{3(1-z)}{u}+\frac{2L_{W}\left( \varrho
_{u}\right) }{u(1+L_{W}\left( \varrho _{u}\right) )}\right]
A_{x_{1}}^{\prime }\left( u\right)  \notag \\
+\frac{l^{2z+2}u^{2z-2}}{r_{+}^{4z}\mathcal{F}(u)^{2}}\left( \omega ^{2}-%
\frac{4q\beta b^{2z-2}\mathcal{F}(u)}{l^{z+1}}\sqrt{L_{W}\left( \varrho
_{u}\right) }\right) A_{x_{1}}\left( u\right) =0.  \label{E31}
\end{gather}%
Near the $u=0$ boundary, Eq. (\ref{E31}) reduces approximately to Eq. (\ref%
{nearbound}) with the solution (\ref{nearboundsol}). Performing the
procedure of appendix \ref{app1}, the holographic conductivity can be
obtained as%
\begin{equation}
\sigma =\frac{(3z-2)r_{+}^{6z-4}A^{1}}{4\pi i\omega b^{2z-2}l^{z+1}A^{0}}%
\left. \exp \left[ -\frac{1}{2}\left( \frac{l^{z}\omega
A^{0}r_{+}^{2z-6}u^{3-z}}{\beta b^{2z-2}}\right) ^{2}\right] \right\vert
_{u=0}.  \label{holcond}
\end{equation}%
As one can see from (\ref{holcond}), the behaviour of conductivity is
different for different ranges of $z$

\begin{equation}
\sigma =\left\{ 
\begin{array}{ll}
\frac{(3z-2)r_{+}^{6z-4}A^{1}}{4\pi i\omega b^{2z-2}l^{z+1}A^{0}}, & \text{%
for }z<3 \\ 
\frac{(3z-2)r_{+}^{6z-4}A^{1}}{4\pi i\omega b^{2z-2}l^{z+1}A^{0}}\exp \left[
-\frac{1}{2}\left( \frac{l^{z}\omega A^{0}r_{+}^{2z-6}}{\beta b^{2z-2}}%
\right) ^{2}\right] , & \text{for }z=3 \\ 
0, & \text{for }z>3%
\end{array}%
\right.
\end{equation}

\subsubsection{Behavior of the conductivity and experimental results \label%
{num}}

\begin{figure}[t]
\centering{%
\subfigure[]{
   \label{fig11a}\includegraphics[width=.46\textwidth]{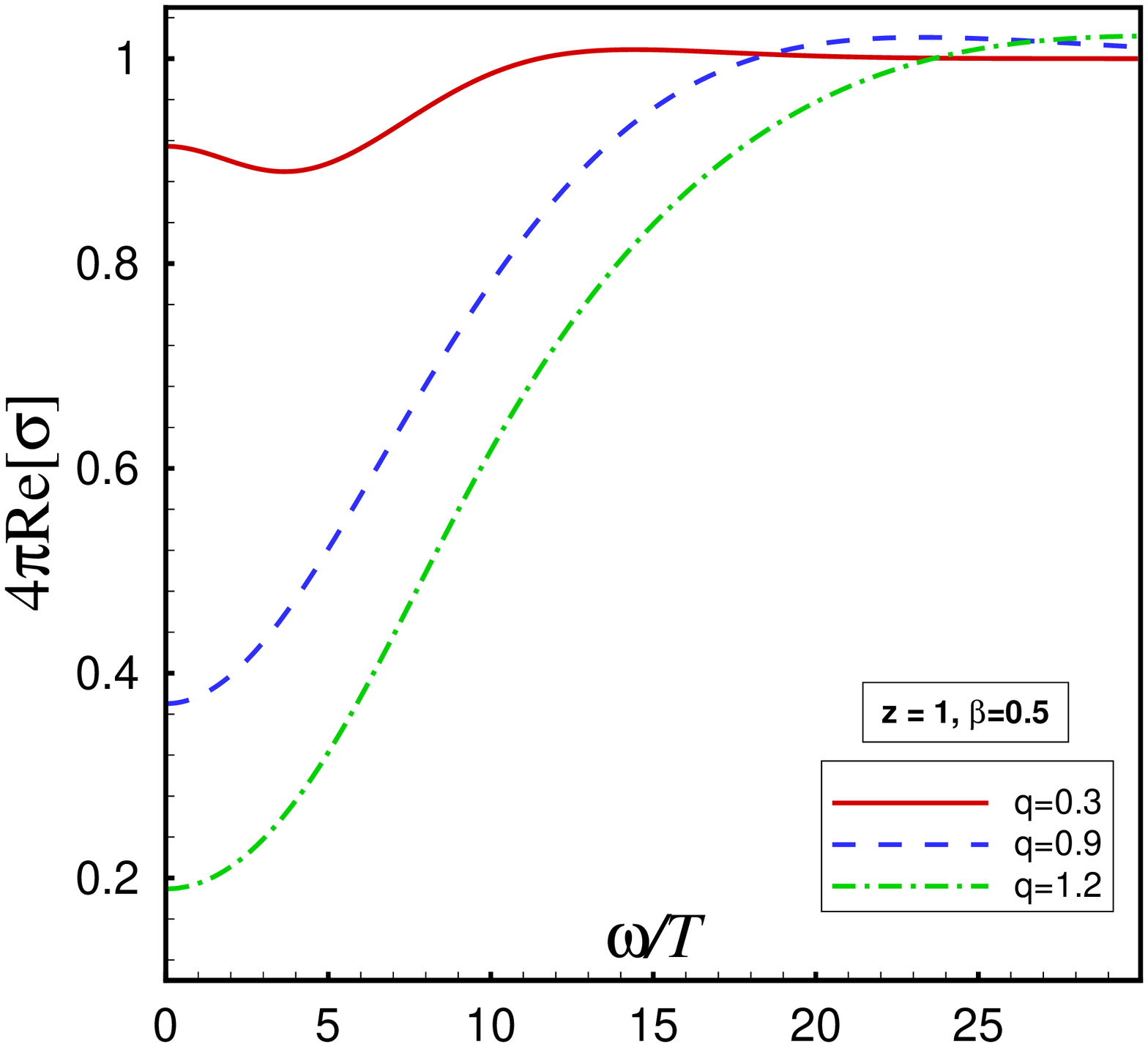}\qquad}} 
\subfigure[]{
   \label{fig11b}\includegraphics[width=.46\textwidth]{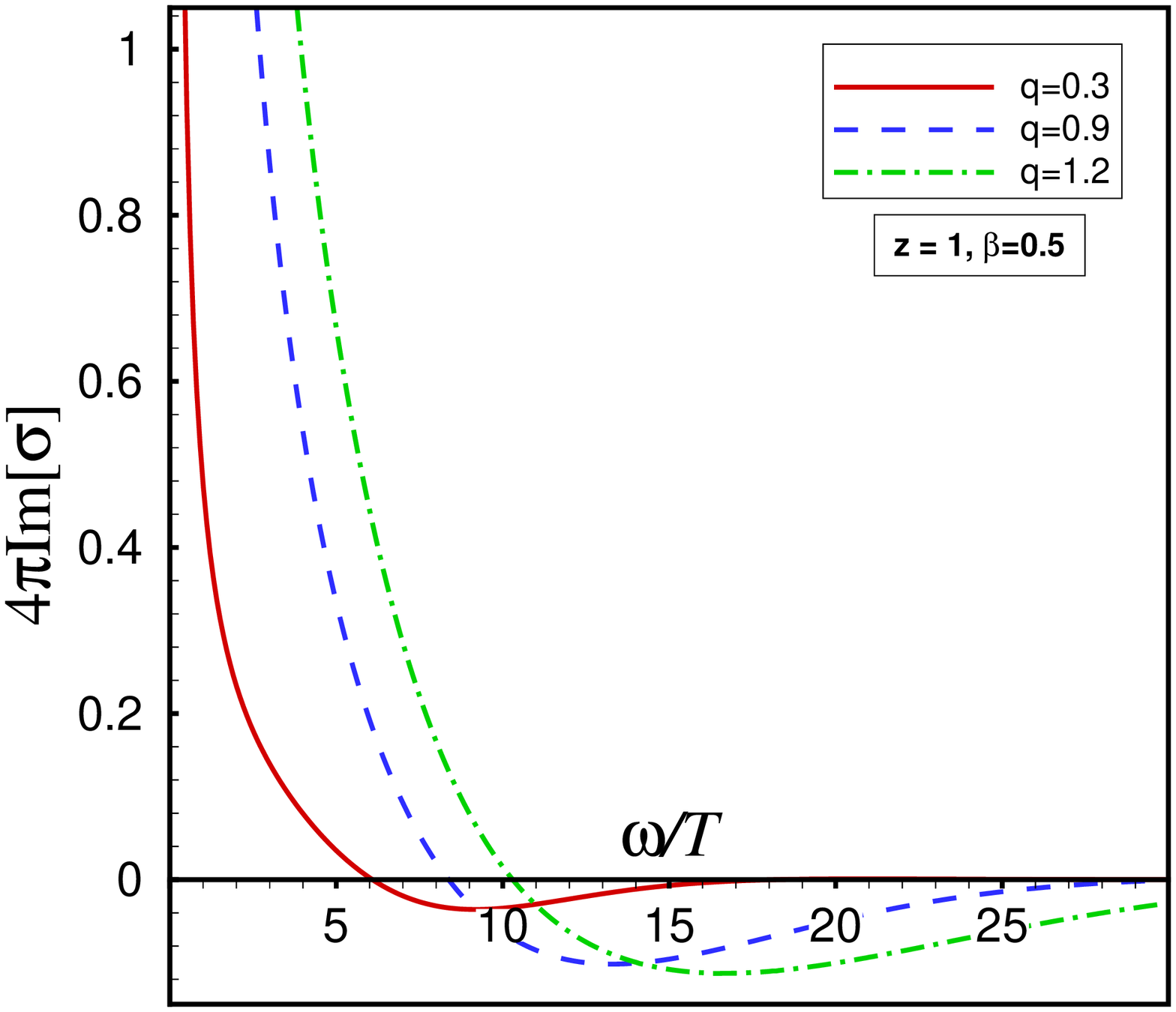}}
\caption{The behaviors of real and imaginary parts of conductivity $\protect%
\sigma $ versus $\protect\omega /T$ for $z=1$ and different values of $q$
with $l=b=r_{+}=1$.}
\label{fig11}
\end{figure}

\begin{figure*}[t]
\centering{%
\subfigure[]{
   \label{fig12a}\includegraphics[width=.46\textwidth]{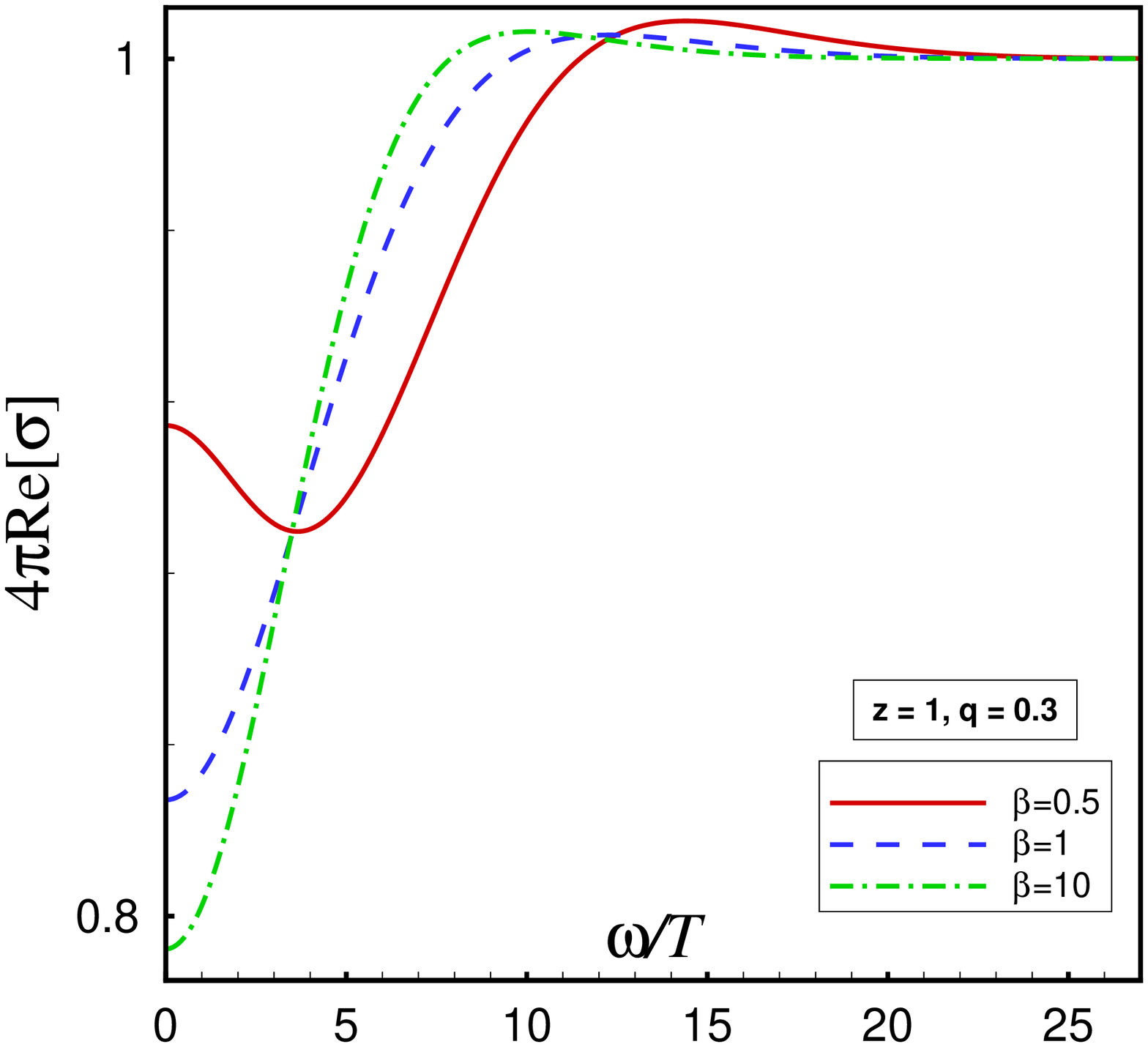}\qquad}} 
\subfigure[]{
   \label{fig12b}\includegraphics[width=.46\textwidth]{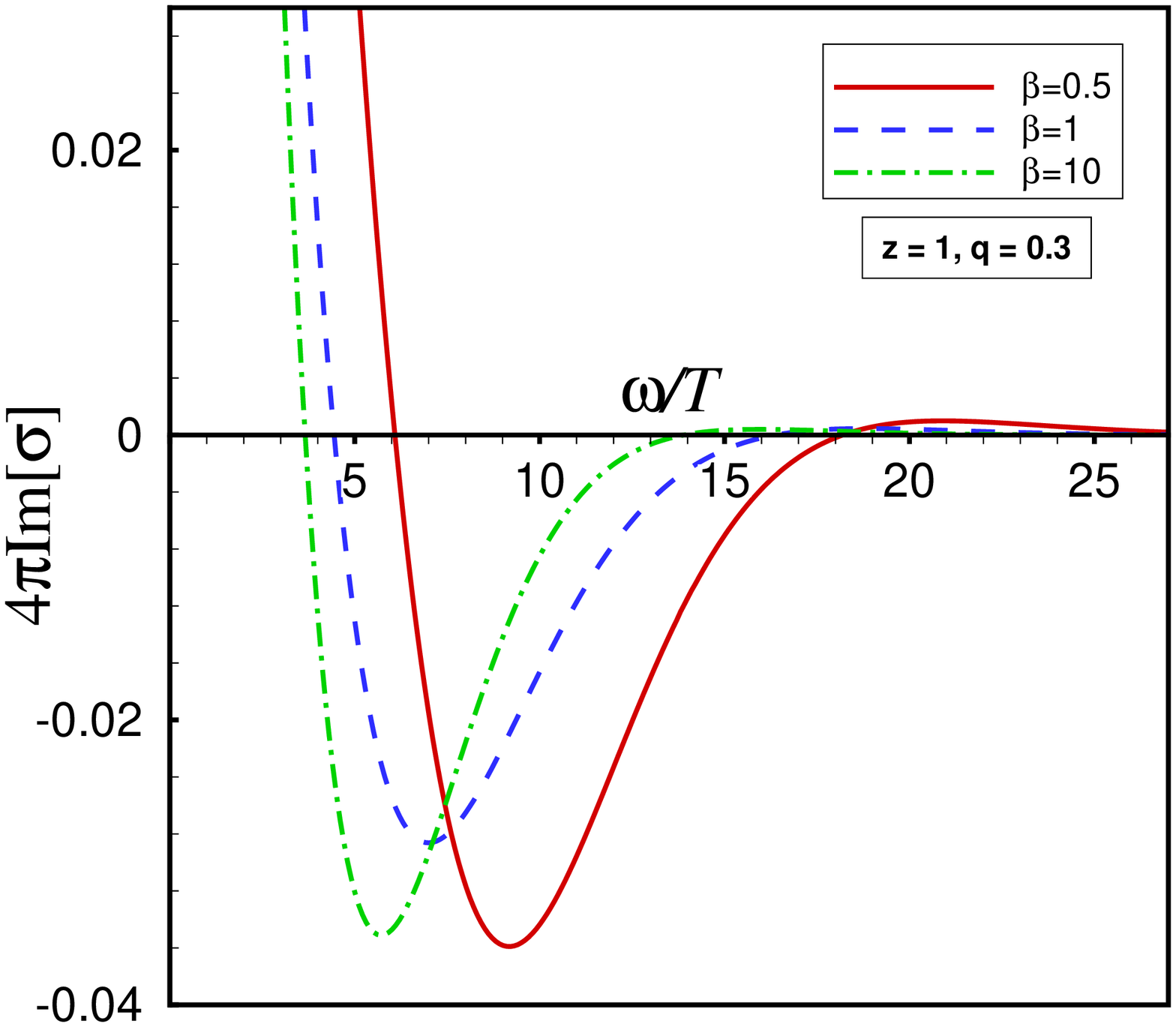}}
\caption{The behaviors of real and imaginary parts of conductivity $\protect%
\sigma $ versus $\protect\omega /T$ for $z=1$ and different values of $%
\protect\beta $ with $l=b=r_{+}=1$.}
\label{fig12}
\end{figure*}

\begin{figure*}[t]
\centering{%
\subfigure[]{
   \label{fig13a}\includegraphics[width=.46\textwidth]{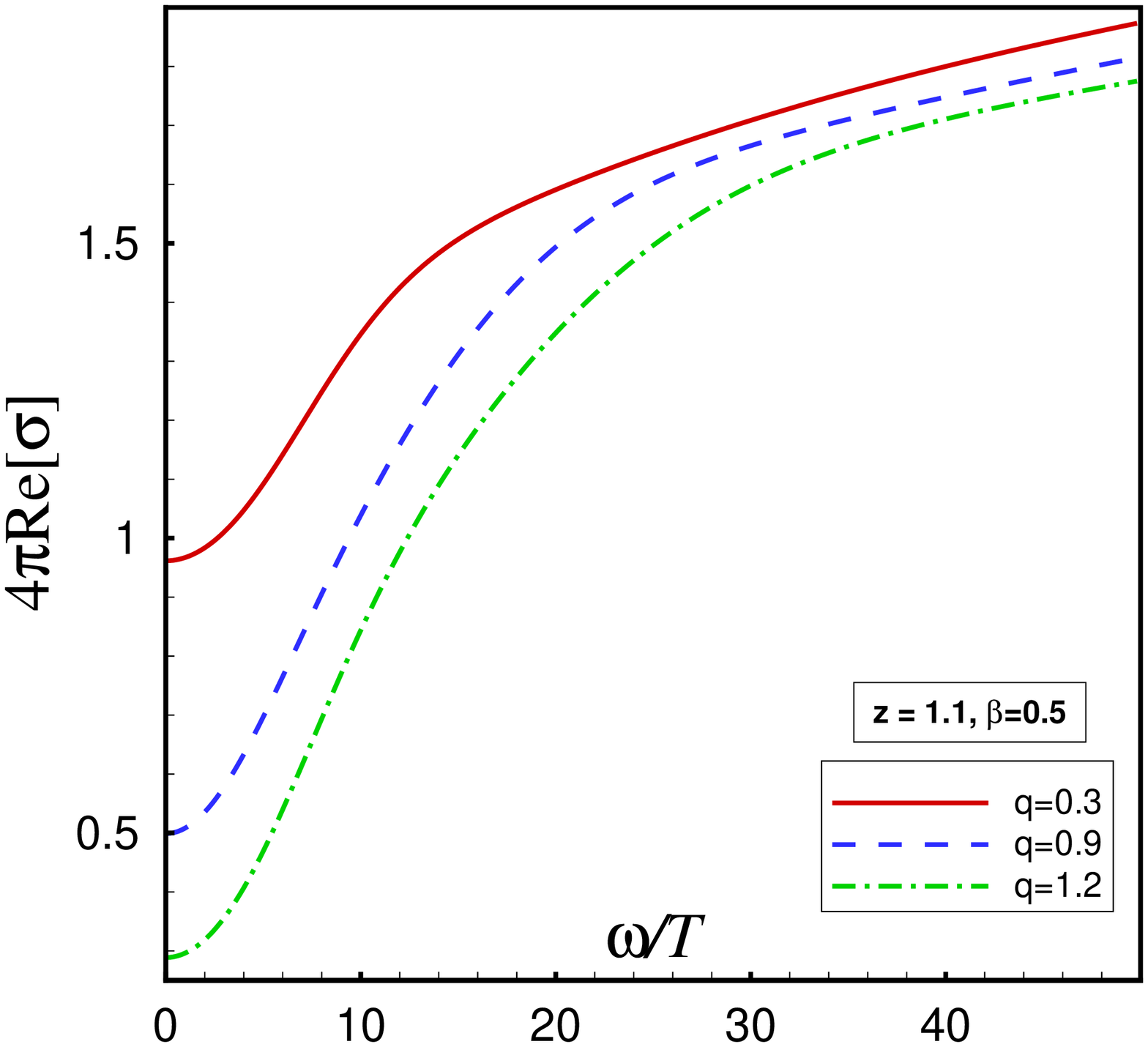}\qquad}} 
\subfigure[]{
   \label{fig13b}\includegraphics[width=.46\textwidth]{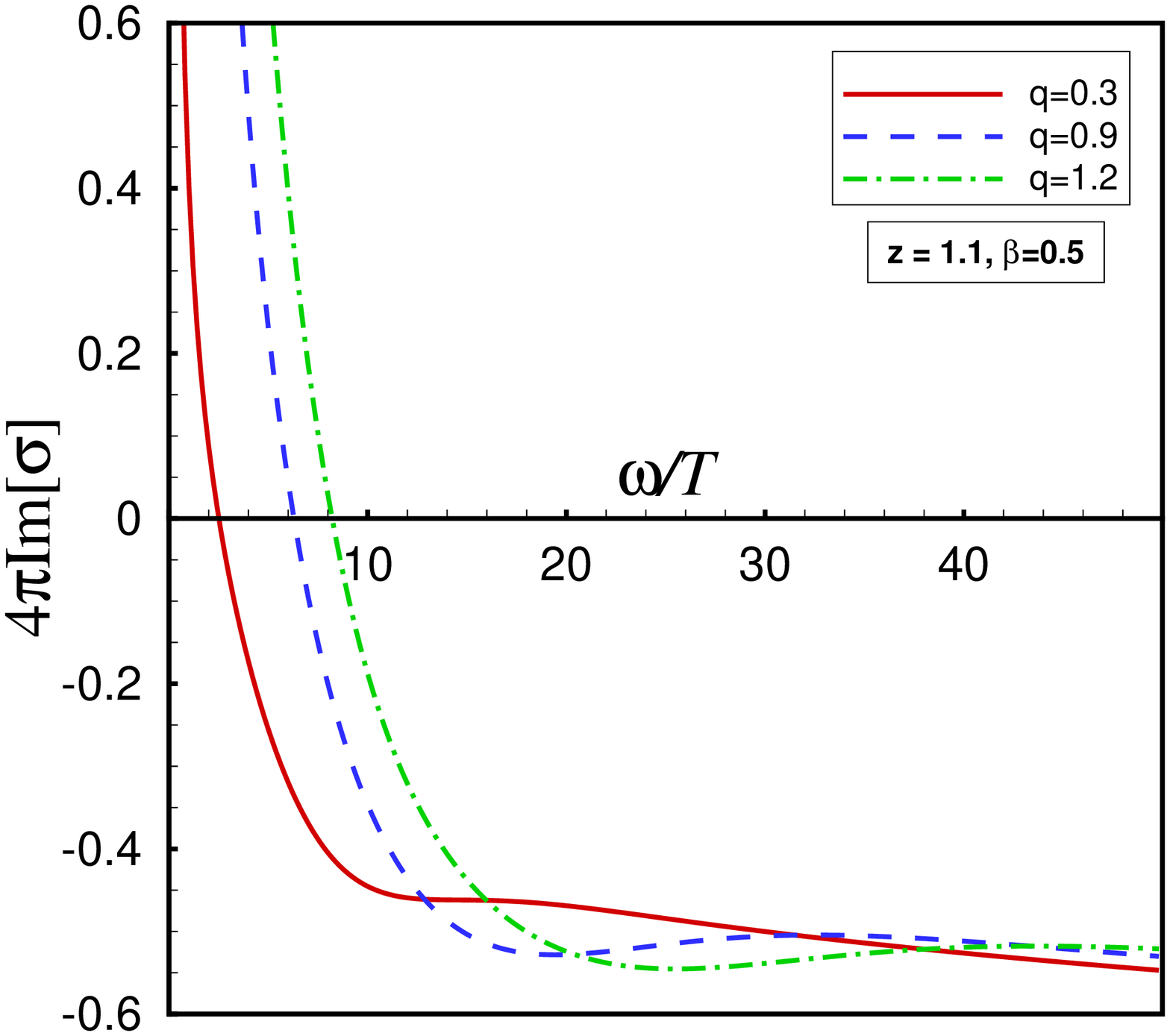}}
\caption{The behaviors of real and imaginary parts of conductivity $\protect%
\sigma $ versus $\protect\omega /T$ for $z=1.1$ and different values of $q$
with $l=b=r_{+}=1$.}
\label{fig13}
\end{figure*}

\begin{figure*}[t]
\centering{%
\subfigure[]{
   \label{fig14a}\includegraphics[width=.46\textwidth]{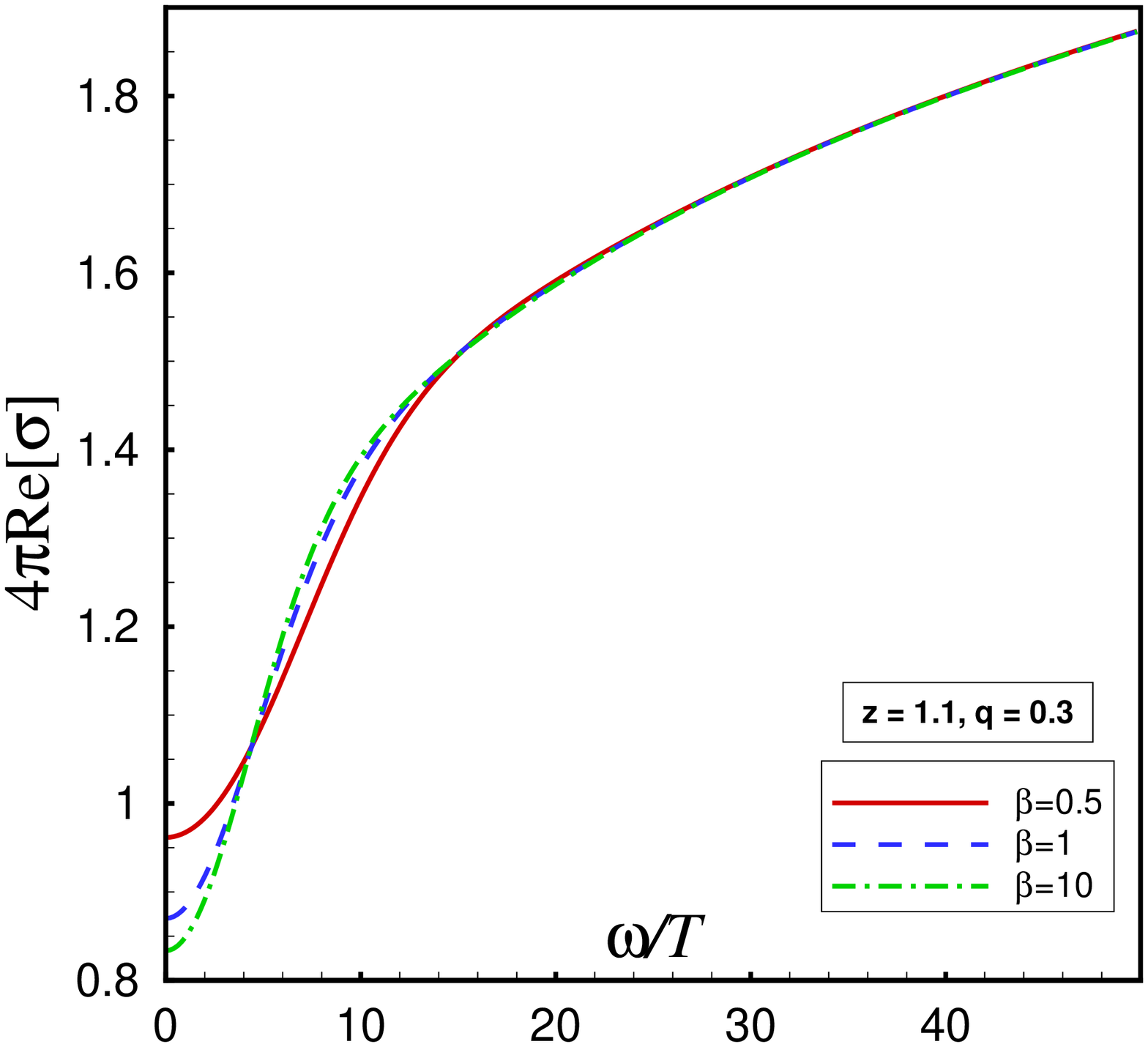}\qquad}} 
\subfigure[]{
   \label{fig14b}\includegraphics[width=.46\textwidth]{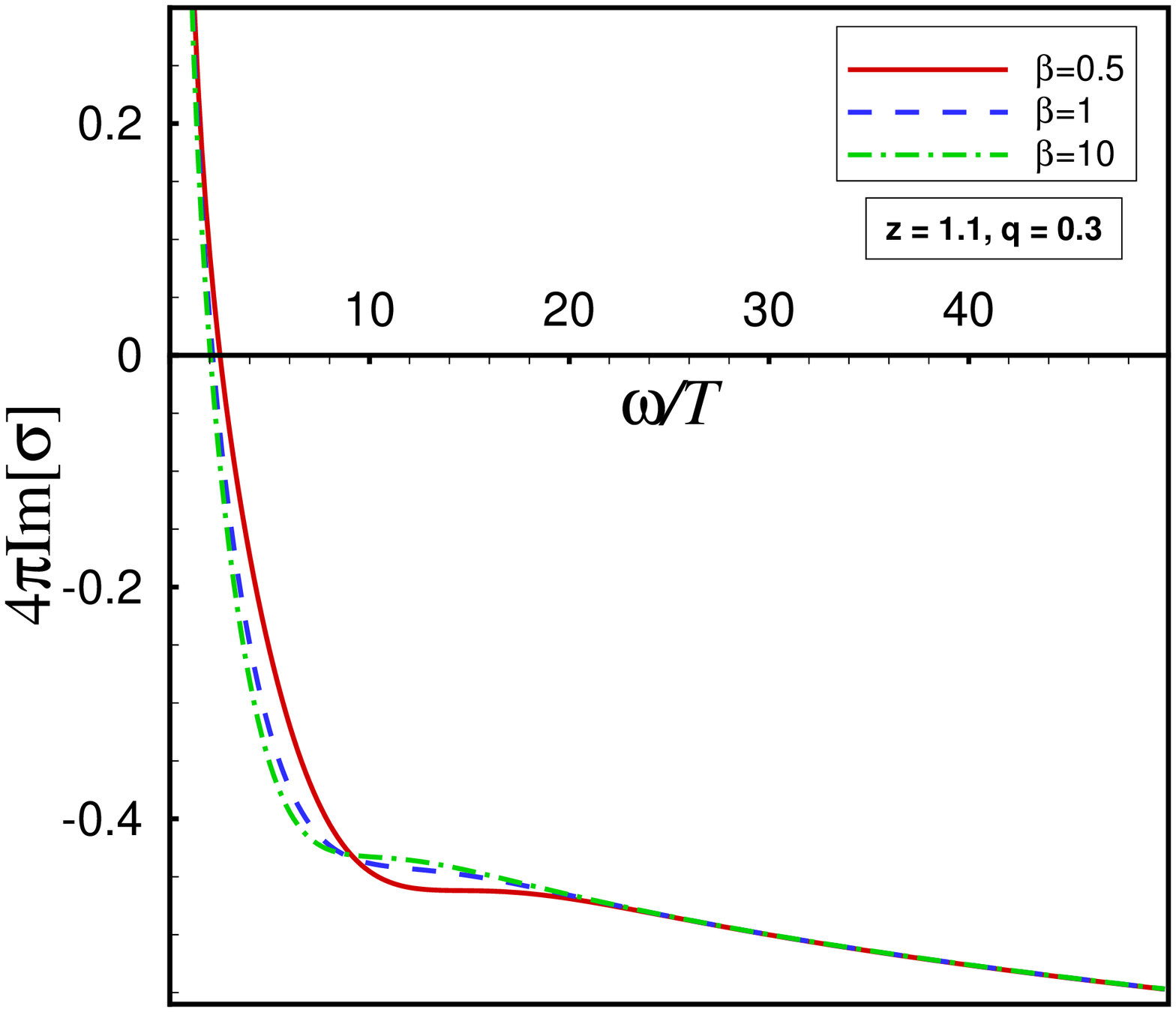}}
\caption{The behaviors of real and imaginary parts of conductivity $\protect%
\sigma $ versus $\protect\omega /T$ for $z=1.1$ and different values of $%
\protect\beta $ with $l=b=r_{+}=1$.}
\label{fig14}
\end{figure*}

In order to illustrate the behavior of the conductivity, we should solve the
decoupled differential equation for $A_{x_{1}}$. By defining%
\begin{equation}
A_{x_{1}}\left( u\right) =\mathcal{F}(u)^{-i4\pi \omega /T}S(u),  \label{Ax}
\end{equation}%
near the horizon, where $S(u)=1+a(u-r_{+})+b(u-r_{+})^{2}+\cdots $, we
remove the oscillations for numerical stability. It is remarkable to note
that at the horizon where $S(u)=1$, $A_{x_{1}}\left( u\right) \propto 
\mathcal{F}(u)^{\pm i4\pi \omega /T}$. However, we choose $\mathcal{F}%
(u)^{-i4\pi \omega /T}$ in (\ref{Ax}) in order to perform ingoing boundary
condition. The coefficients $a$, $b$,$\cdots $ can easily be found by
looking for Taylor series expansion of differential equation (\ref{E3}) at
the horizon. These coefficients are initial values necessary for solving Eq.
(\ref{E3}) numerically to obtain the conductivity.

Figs. \ref{fig11} and \ref{fig12} depict the real and imaginary parts of the
conductivity for $z=1$ in terms of $\omega /T$. It is notable to mention
that the real part of the conductivity is the dissipative part while the
imaginary part is the reactive one. Fig. \ref{fig11} shows that $\sigma
_{DC}\equiv \mathrm{Re}\left[ \sigma \left( \omega =0\right) \right] $ is
greater for smaller values of $q$. The same behavior can be seen for the
nonlinear parameter $\beta $ i.e. $\sigma _{DC}$ increases when $\beta $
decreases. The behaviors of the real and imaginary parts of the conductivity
are in excellent agreement with the experimental results reported in \cite%
{exp1}. These experimental results show the increase in $\mathrm{Re}\left[
\sigma \right] $\ near the zero frequency due to impurities and ionic
lattice. The holographic conductivity computed by employing a gravity dual
model with linear Maxwell electrodynamics cannot produce this behavior \cite%
{hart}. However, as one can see in Figs. \ref{fig11} and \ref{fig12}, for
suitable choices of parameters the nonlinearity of electrodynamics model can
result this behavior for real part of conductivity.

The behaviors of the real and imaginary parts of the conductivity in terms
of $\omega /T$ for $z=1.1$ are illustrated in Figs. \ref{fig13} and \ref%
{fig14}. In this case, $\sigma _{DC}$ changes with respect to $q$ and $\beta 
$ in the similar manner that it does for $z=1$. However, in contrast to the
asymptotic AdS spaces in which the real part of the conductivity tends to a
constant in large frequencies, in this case ($z=1.1$) \textrm{Re}$\left[
\sigma \right] $ grows. Similar behavior for \textrm{Re}$\left[ \sigma %
\right] $ has been reported recently for the optical conductivity of
single-layer graphene induced by mild oxygen plasma exposure \cite{exp2}. It
is remarkable to note that our numerical calculations show that there is no
significant difference between the behaviour of conductivity for $\beta =10$
and larger.

\section{Summary and Concluding Remarks}

In this paper, we constructed a new class of Einstein-dilaton-Lifshitz black
solutions in the presence of exponential nonlinear electrodynamics. Our
solutions respect the Lifshitz symmetry $t\rightarrow \lambda ^{z}t$ and $%
\vec{\mathbf{x}}\rightarrow \lambda \vec{\mathbf{x}}$ at $r$-infinity
boundary where $z(\geq 1)$ is the dynamical critical exponent. The
exponential electrodynamics behaves as Born-Infeld electrodynamics for large
values of the nonlinear parameter $\beta $ where they reduce to linear
Maxwell regime, as expected. It is worth mentioning that, while the
Born-Infeld nonlinear electrodynamics removes divergences in the electric
field and has finite value \textit{near the origin} where $r\rightarrow0$,
the exponential form of the nonlinear electromagnetics does not cancel the
divergency of the electric field exactly at $r = 0$, however, its
singularity is much weaker than Maxwell theory. This is more reasonable
compared to the Born-Infeld case, since \textit{near the origin} where $%
r\rightarrow0$, the electric field of a point-like charged particle should
be an increasing function. The behavior of the electric field of the
exponential nonlinear electrodynamics \textit{near} the origin in the
absence and in the presence of the dilaton field was explicitly shown in
table A of Refs. \cite{hendi2} and \cite{SheyKaz}, respectively. Besides, it
was argued that in applications of the AdS/CFT correspondence to
superconductivity, exponential nonlinear electrodynamics, makes crucial
effects on the condensation as well as the critical temperature of the
superconductor \cite{ZPCJ.NN}. It was also recently observed that, in the
holographic superconductor, the exponential nonlinear electrodynamics can
increase the critical values of the external magnetic field as the
temperature goes to zero \cite{SheAb}.

We considered topological black holes with zero ($k=0$), positive ($k=1$)
and negative ($k=-1$) horizon curvatures, but the reality of charge of
asymptotic Lifshitz supporting Maxwell matter field imposes that there is no
allowed solution with negative horizon curvature except for asymptotic AdS
case ($z=1$). Next, we calculated the conserved and thermodynamical
quantities namely mass, charge, electric potential, entropy and temperature.
We obtained the mass in terms of the extensive quantities entropy $s$ and
charge $Q$ i.e. $M\left( s,Q\right) $ and checked that the intensive
quantities temperature $T$ and electric potential $U$ calculated from it
match with those obtained from the geometry of the black hole. Thus, the
first law of thermodynamics is satisfied.

Subsequently, we turned out to study thermal stability of the obtained
solutions. We showed that black brane solutions ($k=0$) are always thermally
stable in canonical ensemble. In grand-canonical ensemble black branes with
horizon radius $r_{+}$ larger than $r_{+\min }$ are stable while there is a
maximum $z$ ($\beta $) that black branes are unstable for values greater
than it. We observed that black holes ($k=1$) encounter the Hawking-Page
phase transition between small and large black holes in both canonical and
grand-canonical ensembles. For $z\geq 2$, our black holes are always stable
in canonical ensemble. For other values of $z$ in canonical ensemble as well
as grand-canonical ensemble, there is a minimum for dynamical critical
exponent that black holes are stable for values greater than it. Moreover,
black holes are stable for nonlinear parameters greater than $\beta _{\min }$
in canonical ensemble while they are unstable for values greater than $\beta
_{\max }$ in grand-canonical ensemble.

Afterward, we performed the gauge/gravity duality to compute the ratio of
shear viscosity to entropy for a three-dimensional hydrodynamic system by
using the pole method. We obtained the well-known $1/4\pi $ result which
shows that even the non-AdS symmetry of hydrodynamic system cannot affect
the value of viscosity. Finally, we turned to study the behavior of the
holographic conductivity for two-dimensional systems. We examined the issue
for both linear Maxwell and nonlinear exponential electrodynamics. Our
investigations revealed that the effect of nonlinearity is vanishing the
conductivity for $z>3$. We depicted the behaviors of real and imaginary
parts of conductivity for asymptotic AdS ($z=1$) and Lifshitz cases and
pointed out some two-dimensional graphene systems taken under consideration
experimentally which have behaviors resemble our numerical results.

\acknowledgments{We thank from the Research Council of Shiraz
University. This work has been supported financially by Research
Institute for Astronomy \& Astrophysics of Maragha (RIAAM), Iran.}

\appendix

\section{A Brief Review on Gauge-Gravity Duality Basics}

\label{app1} Here we are going to present a brief review on the idea of
gauge-gravity duality (reader can see for instance \cite{hart} and \cite%
{tong} for more details). Let us start with QFT. The generating function in
QFT $Z_{\mathrm{QFT}}$ has a central role. For instance the expectation
value of an operator $\mathcal{O}$ sourced by $\phi _{0}$ is given by%
\begin{equation}
\left\langle \mathcal{O}\right\rangle =\frac{1}{Z_{\mathrm{QFT}}\left[ \phi
_{0}\right] }\frac{\partial Z_{\mathrm{QFT}}\left[ \phi _{0}\right] }{%
i\partial \phi _{0}}=\frac{\partial \ln \left( Z_{\mathrm{QFT}}\left[ \phi
_{0}\right] \right) }{i\partial \phi _{0}}.
\end{equation}%
For a strongly interacting field theory, it is hard to compute the
generating function $Z_{\mathrm{QFT}}$ and here is the position where
gauge-gravity duality helps us. Consider a bulk that our field theory lives
on its $u=0$ boundary. The function $\phi _{0}\left( x\right) $ on the
boundary becomes a field $\phi \left( x,u\right) $ governed by an equation
of motion in the bulk so that $\phi \left( x,u\right) \rightarrow \phi
_{0}\left( x\right) $ as one approaches the boundary $u\rightarrow 0$. The
fundamental formula of holography is GKPW formula (Gubser, Klebanov,
Polyakov \cite{GKP} and Witten \cite{Witt}) that its classical gravity limit
is%
\begin{equation}
Z_{\mathrm{QFT}}\left[ \phi _{0}\right] =e^{iS_{\mathrm{bulk}}},
\label{GKPW}
\end{equation}%
where $S_{\mathrm{bulk}}$ is the action calculated by using solutions of
bulk equations of motion obtained subject to the requirement that $\phi
\rightarrow \phi _{0}$ on the boundary. Using (\ref{GKPW}), one can find that%
\begin{equation}
\left\langle \mathcal{O}\right\rangle =\frac{\partial S_{\mathrm{bulk}}}{%
\partial \phi _{0}}.  \label{expval}
\end{equation}%
In order to find a more convenient alternative for (\ref{expval}), we pause
here to remind the Hamilton-Jacobi theory. Varying the action of a point
particle with position $x$, one receives%
\begin{equation}
\delta S_{\mathrm{particle}}=\int_{t_{i}}^{t_{f}}dt\left[ \frac{\partial L}{%
\partial x}-\frac{d}{dt}\left( \frac{\partial L}{\partial \dot{x}}\right) %
\right] \delta x+\left[ \frac{\partial L}{\partial \dot{x}}\delta x\right]
_{t_{i}}^{t_{f}}.
\end{equation}%
Considering the on-shell action and supposing that initial position of the
particle is fixed ($\delta x\left( t_{i}\right) =0$) while the final
position is varying ($\delta x\left( t_{f}\right) \neq 0$), one obtains%
\begin{equation}
\frac{\partial S}{\partial x_{f}}=\left. \frac{\partial L}{\partial \dot{x}}%
\right\vert _{t_{f}}=p_{f}.
\end{equation}%
We can generalize the above result to our case and rewrite (\ref{expval}) as%
\begin{equation}
\left\langle \mathcal{O}\right\rangle =\left. \frac{\partial \mathcal{L}}{%
\partial \left( \partial _{u}\phi \right) }\right\vert _{u=0}.
\end{equation}%
Therefore, we could obtain the expectation value of the operator $\mathcal{O}
$ without calculating $Z_{\mathrm{QFT}}$. Note that the expectation value
was a simple example to present the basic ideas of gauge-gravity duality.
Another remark to note is that the operator $\mathcal{O}$ and its source $%
\phi _{0}$ which are scalars can be extended to vectors or tensors. For
instance the source for $J^{i}$ is $A_{0}^{i}$ and the source for $T^{ij}$
is $g_{0}^{ij}$. As an example the expectation value of $J_{i}$ sourced by
the perturbation $\delta A_{0i}=A_{0i}e^{-i\omega t}$ is%
\begin{equation}
\left\langle J_{i}\right\rangle =\left. \frac{\partial \mathcal{L}}{\partial
\left( \partial _{u}\delta A_{i}\right) }\right\vert _{u=0}.  \label{cuden}
\end{equation}%
Then, with (\ref{cuden}) in hand we can obtain the conductivity via the
formula $\sigma =\left\langle J_{i}\right\rangle /E_{i}$.

\end{document}